\documentclass[12pt]{article}

\usepackage[numbers,sort&compress]{natbib}
\usepackage{graphicx}
\usepackage{amsmath}
\usepackage{amssymb}
\usepackage{hyperref}

\newcommand{\beq}{\begin{equation}}
\newcommand{\eeq}{\end{equation}}
\newcommand{\beqs}{\begin{eqnarray}}
\newcommand{\eeqs}{\end{eqnarray}}
\newcommand{\Tr}{\ensuremath{\mathop{\mathrm{Tr}}}}
\newcommand{\sign}{\ensuremath{\mathop{\mathrm{sign}}}}

\def\ni{\noindent}
\def\be{\begin{equation}}
\def\ee{\end{equation}}
\def\bea{\begin{eqnarray}}
\def\eea{\end{eqnarray}}
\def\bsp{\be\begin{split}}

\def\ra{\rangle}
\def\dag{\dagger}

\def\lr{\leftrightarrow}

\def\da{\dot{\alpha}}
\def\db{\dot{\beta}}
\def\dg{\dot{\gamma}}
\def\dd{\dot{\delta}}

\def\G{\Gamma}
\def\D{\Delta}

\def\a{\alpha}
\def\b{\beta}
\def\g{\gamma}

\def\d{\delta}
\def\e{\epsilon}
\def\m{\mu}

\def\n{\nu}
\def\s{\sigma}
\def\r{\rho}
\def\l{\lambda}
\def\t{\tau}
\def\o{\omega}
\def\O{\Omega}
\def\T{\theta}

\def\p{\partial}

\def\lr{\leftrightarrow}

\def\bR {\mathbb{R}}
\def\bZ {\mathbb{Z}}

\def\w{\wedge}

\makeatletter

\newcommand{\Rmnum}[1]{\expandafter\@slowromancap\romannumeral #1@}
\makeatother

\renewcommand{\title}[1]{\vbox{\center\LARGE{#1}}\vspace{5mm}}
\renewcommand{\author}[1]{\vbox{\center\large{#1}}\vspace{5mm}}
\newcommand{\address}[1]{\vbox{\center\em#1}}
\newcommand{\email}[1]{\vbox{\center\tt#1}\vspace{5mm}}

\begin{document}
\bibliographystyle{utphys}

\begin{titlepage}
\hfill {\tt HU-EP-10/15}\\
\title{\vspace{1.0in} {\bf SU(2$|$2) for Theories with Sixteen Supercharges
at Weak and Strong Coupling}}

\author{Abhishek Agarwal$^1$ and Donovan Young$^2$}

\address{$^1$American Physical Society\\
  1 Research Road\\
  Ridge, NY 11961, USA\\
  \vspace{.5cm}
  $^2$Humboldt-Universit\"at zu Berlin, Institut f\"ur Physik,\\
  Newtonstra\ss e 15, D-12489 Berlin, Germany }

\email{$^1$abhishek@ridge.aps.org, $^2$dyoung@physik.hu-berlin.de}

\abstract{\ni We consider the dimensional reductions of ${\cal N}=4$
  Supersymmetric Yang-Mills theory on $\bR\times S^3$ to the
  three-dimensional theory on $\bR\times S^2$, the orbifolded theory
  on $\bR\times S^3/\bZ_k$, and the plane-wave matrix model. With
  explicit emphasis on the three-dimensional theory, we demonstrate
  the realization of the $SU(2|3)$ algebra in a radial Hamiltonian
  framework. Using this structure we constrain the form of the spin
  chains, their S-matrices, and the corresponding one- and two-loop
  Hamiltonian of the three dimensional theory and find putative signs
  of integrability up to the two loop order. The string duals of these theories admit the IIA
  plane-wave geometry as their Penrose limit. Using known results for
  strings quantized on this background, we explicitly construct the
  strong-coupling dual extended $SU(2|2)$ algebra and discuss its
  implications for the gauge theories.}

\end{titlepage}

\tableofcontents
\section{Introduction and summary}

Mass-deformed Lie superalgebras continue to play an important r\^ole in
deepening our understanding of the gauge-gravity
correspondence\footnote{Because we will be dealing with string theory
  backgrounds which do not contain an $AdS$ subspace, and
  non-conformal dual field theories, we will avoid use of the more
  colloquial term ``$AdS/CFT$'' in this context.}. The algebras in
question are generically of the form
\be \{Q^\dag, Q\} =
P + mR, 
\ee 
where $R$ is typically some combination of spacetime and R-symmetry
rotation generators and $m$ denotes a basic ``mass-gap'' in the
spectrum of observables constrained by the supersymmetry algebra.
Perhaps the most dramatic manifestation of this algebraic structure is
the integrability of the large-$N$ dilatation operator of
$\mathcal{N}=4$ supersymmetric Yang-Mills theory ($\mathcal{N}=4$
SYM), where the ``mass-gap'' is the gap in the spectrum of conformal
dimensions $\D$ of the operators of the theory, for which $\D\geq 2$.
One of the basic building blocks on which much of the machinery that
renders the four dimensional superconformal theory integrable rests is
the closed $SU(2|3)$ sector. The matter fields in this sector comprise
two Weyl fermions $\psi_\alpha, $ $\alpha = 1,2$, transforming under
an $SU(2)$ of the Lorentz group $L$, and three complex scalars
$X,Y,Z$, transforming under an $SU(3) \in SU(4) \sim SO(6)$ of the
R-Symmetry group $R$. The relevant supercharges $S^\alpha _a$,
$Q^b_\beta$, satisfy
\be
\{S^\alpha _a, Q^b_\beta\} = \delta ^b_a\delta ^\alpha _\beta D +
\delta ^b_a L^\alpha _\beta + \delta ^\alpha _\beta R^b_a ,
\ee
where $a,b$ are $SU(2)$ indices corresponding to an $SU(2)$ contained
in the R-symmetry group.  This is nothing but a subalgebra of the
full four dimensional superconformal algebra and $S,Q$ are a subset of
the superconformal and supersymmetry generators respectively
\cite{Beisert:2003ys}.  $D$ is the dilatation operator, which is
realized as a quantum spin chain in the large-$N$ limit. The
ferromagnetic ground state of the chain is spanned by the chiral
primary operators $\Tr(Z^J)$. The excitations/magnons transform
under the residual $SU(2|2)$ symmetry, which has profound
consequences.  For instance, the $S$-matrix of the spin chain, and its
dispersion relation are both severely constrained to all orders in
perturbation theory by this symmetry. The $SU(2|2)$ invariance also
constrains the $S$-matrix to satisfy the Yang-Baxter equations
\cite{Beisert:2005tm, Beisert:2006qh}. It is, of course, natural to
ask if these compelling consequences of the underlying (mass-deformed)
supersymmetry algebra have any repercussions for theories other than
$\mathcal{N} =4$ SYM.

Recent developments point out that the $SU(2|2)$ symmetry plays a
greater r\^ole in the gauge-gravity correspondence than was previously
realized. For example it appears in the studies of $\mathcal{N}=6$
supersymmetric Chern-Simons (SCS) theories in two different contexts.
The $\mathcal{N}=6$ conformal models appear to posses an integrable
dilatation operator in the large-$N$ limit \cite{Gaiotto:2008cg,
  Minahan:2008hf}. The dilatation operator is part of the center of a
$SU(2|3)$ algebra just as in the case of the four dimensional
superconformal theory mentioned above.  A generalization of the
$\mathcal{N} =6$ SCS models can be obtained by adding appropriate mass
terms to the matter fields. These mass-deformed models can be
engineered to preserve $4\leq \mathcal{N} \leq 8$ supersymmetry at the
expense of conformal invariance. For such models $SU(2|2)$ algebras
also arise as part of the spacetime supersymmetry, and they can be
used to obtain all-loop results for the spacetime $S$-matrices of the
massive theories \cite{Agarwal:2008pu}.  These results clearly suggest
that the search for other natural habitats for mass-deformed
supersymmetry algebras and their consequences for the gauge/gravity
duality are worthwhile endeavors.

A natural set of theories to investigate in this regard are the
dimensional reductions of $\mathcal{N}=4$ SYM on $\mathbb{R}\times
S^3$. The $S^3$ can obviously be identified with $SU(2)$. Discarding
the dependence of the degrees of freedom of the four dimensional
theory on $\bZ_k $, $U(1)$, and\footnote{Both $Z_k$ and $U(1)$ $\in$
  $SU(2) \sim S^3$.} $SU(2)$ produces 16 supercharge theories with
massive spectra on $\mathbb{R}\times S^3/\bZ_k$, $\mathbb{R}\times
S^2$, and $\mathbb{R}$ respectively \cite{Lin:2005nh}. The theory on
$\mathbb{R}$ is nothing but the plane-wave matrix model (PWMM)
\cite{Kim:2003rza} while the other dimensional reductions result in
gauge theories with massive spectra\footnote{The mass for the scalars
  arises form the conformal coupling of the original four dimensional
  theory to the radius of $S^3$. The masses of the gluons are simply a
  consequence of the lack of a zero mode for vector fields on $S^2$
  and $S^3$.}. Concrete proposals for the dual string theories
corresponding to these dimensionally reduced models were also
enunciated in \cite{Lin:2005nh}. Since the $SU(2|3)$ symmetry of the
four dimensional gauge theory is preserved by these dimensional
reductions it is imperative to try and uncover its consequences in the
gauge/string dualities tying the lower dimensional non-conformal
theories to string theories in non-$AdS$ type backgrounds.  This line
of investigation also has the advantage of being a valuable probe for
the utility and robustness of the gauge-gravity conjecture for massive
and non-conformal gauge theories and their dual string theory
backgrounds. In much of the analysis that we perform, we concentrate
on the three dimensional gauge theory and its string dual, while
commenting on the generalizations of our results to the orbifold
theory and the plane wave matrix model where we can do so.

Since much of our intuition about the use of mass-deformed
superalgebras derives from studies of the dilatation operator of
$\mathcal{N}=4$ SYM on $\mathbb{R}^4$, it is instructive to recall
how the conformal transformation mapping the theory to
$\mathbb{R}\times S^3$ affects various elements of the superconformal
algebra. In the flat background the superconformal algebra takes on
the following heuristic form \bsp\label{superconf}
&\{Q, Q\} = P,\\
&\{S, S\} = K,\\
&\{S, Q\}  = D + R + L,
\end{split}
\ee where $K$ is the super-boost generator. Going from the flat space
to $S^3$ amounts to radial quantization of the conformal theory. Under
this quantization scheme, the generators map as follows
\cite{Kinney:2005ej}: \bsp\label{conf} Q &\rightarrow Q,~~S
\rightarrow Q^\dag, ~~ D \rightarrow H,~~ t \rightarrow r.
\end{split}
\ee The relation between $Q^\dag$ and $S$ is a consequence of the
natural hermiticity properties endowed on the physical states upon the
conformal transformation \cite{Kinney:2005ej}. The last two relations
in (\ref{conf}) simply imply that dilatations are mapped to radial
scalings in the scheme of radial quantization, with the Hamiltonian
assuming the r\^ole of $D$ \cite{Fubini:1972mf}. These identifications
enable us to recover a mass-deformed algebra (the analog of the last
equation in (\ref{superconf})) even in the absence of conformal
symmetries, as part of the supersymmetry algebra for the gauge theory
on $\mathbb{R}\times S^3$. The SUSY algebra takes on the following
generic form
\be\label{radialsusy} \{Q^\dag, Q\} = H + \mu R + \mu L ,\ee
where $\m\sim(\text{radius of}~S^3)^{-1}$.  
It is important to note that the same basic algebraic structure is
also valid for 16 supercharge theories on $\mathbb{R}\times S^3/\bZ_k$,
$\mathbb{R}\times S^2$ and $\mathbb{R}$, even though, for these
theories there is no natural sense in which they are equivalent to
their massless counterparts.  Recasting the details of the techniques applied in computing 
the spectrum of the dilatation operator of the four dimensional gauge
theory and its string dual in the radial quantization scheme
should then allow us to study and solve for the physical spectrum of
these non-conformal gauge theories. 

As mentioned above, this proposal has the potential of
translating into a non-trivial test of the gauge-gravity
conjecture in light of the existence of bona fide string duals for the
non-conformal lower dimensional massive gauge theories. These are the
Lin-Maldacena geometries \cite{Lin:2005nh}, which we review in section
\ref{sec:string}. Quantizing the superstring in these backgrounds has
only been successful in the plane-wave limit, yielding the same IIA
plane-wave geometry for any of the Lin-Maldacena
backgrounds\footnote{There is a caveat here concerning the vacuum of
  the gauge theory under consideration. The plane-wave geometry is
  only valid for ``well-spaced'' vacua. The meaning of ``well spaced''
  is described in section \ref{sec:string}.\label{foot:cav} }.
Happily, the plane-wave limit is sufficient for uncovering the
string-dual manifestation of the $SU(2|2)$ algebra we find in the
gauge theories. The crucial element is the derivation of the central
charges. For the full superstring on $AdS_5\times S^5$, the authors of
\cite{Arutyunov:2006ak} showed insightfully that the central extension
of the algebra follows from a relaxation of the level-matching
condition, the natural dual of the length-changing action in the gauge
theory. We show that the very same mechanism is at play for the
plane-wave limit of the Lin-Maldacena geometries, and this allows us a
full exhibition of the $SU(2|2)$ algebra in that limit. Under the
assumption that the algebra persists beyond the plane-wave limit, we
are able to discuss, on a qualitative level, the finite-size
corrections, worldsheet scattering matrix, spinning string and giant
magnon solutions associated with these string sigma models.

This results in this paper are organized as follows. In section
\ref{sec:su22} we explicitly construct the $SU(2|3)$ algebra in SYM on
$\bR\times S^3$ and on $\bR\times S^2$. In sections \ref{su22-gauge}
and \ref{weak} we obtain the dispersion relation for SYM on $\bR\times
S^2$, and constrain the form of the one and two-loop effective
Hamiltonian. We also present evidence of integrability at the two loop
level for the three dimensional theory. In section \ref{sec:intS} we
show the natural generalization of the $SU(2|2)$ $S$-matrix from SYM
on $\bR\times S^3$ to SYM on $\bR\times S^2$.  Our results concerning
the universal form of the dispersion relation and S-matrix are
expected to remain valid to all loop orders. However, as far as the
explicit forms of the effective Hamiltonians and statements about
integrability are concerned, the present gauge theoretic analysis is
restricted to the two-loop order.  In section \ref{sec:extend} we
discuss the extension of our results to the PWMM and SYM on $\bR\times
S^3/\bZ_k$. In section \ref{sec:string} we continue the analysis to
the leading order at strong coupling using the dual string theory. We
discuss the Lin-Maldacena geometries and review the quantization of
the string on the IIA plane-wave. We then derive the $SU(2|2)$ algebra
in the plane-wave setting and discuss implications and further
directions.  Finally, we end with a discussion in section
\ref{sec:final}.


\section{$SU(2|2)$ in SYM on $\bR\times S^3$ and $\bR\times S^2$}
\label{sec:su22}

In this section, we isolate the closed $SU(2|3)$ sector in the sixteen
supercharge Yang-Mills theories on $\bR\times S^3$ and
$\mathbb{R}\times S^2$. We start with the four dimensional theory in
radial quantization and present the details of the emergence of the
algebraic structure (\ref{radialsusy}) in a Hamiltonian picture. The
analysis of the four dimensional model also allows us to calibrate and
verify our results against known results for the dilatation operator
for the gauge theory in flat background geometries.  We find it
convenient to use the conventions used in \cite{Ishiki:2006yr}, and use
the action 
\bsp 
S = \frac{1}{g^2} \Tr \int_{\mathbb{R}\times S^3}
\Biggl[ &-\frac{1}{4} F_{ab}^2 -
\frac{1}{2} D_a X_{m} D^a X^m -\frac{1}{2}\left(\frac{\mu}{2}\right)^2
X_mX^m -\frac{ i}{2} \bar \l \Gamma^a D_a \l\\
&+\frac{1}{4}[X_m,X_n]^2 -\frac{1}{2}\bar \l \Gamma^m[X_m,\l] \Biggr],
\end{split}\label{actions3}
\ee
where $m,n=1,\ldots,6$ are $SO(6)$ indices and $a,b=0,\ldots,3$ are
spacetime indices. It is understood that we have normalized the
radius of $S^3$ to be $2/\mu$, such that the volume is given by
$2\pi^2(2/{\mu})^3$. In other words,
\be
\int_{\mathbb{R}\times S^3} \equiv
\left(\frac{1}{8}\right)\left(\frac{2}{\mu}\right)^3\int dt\int_0^\pi
\sin\theta d\theta \int_0^{2\pi}d\phi\int_0^{4\pi}d\psi. 
\ee 
We can always adjust the radius to be any other number by
correspondingly scaling the coefficient of the ``mass-term'' for the
scalars. $\Gamma^M = (\gamma^a \otimes I_8,\G^m)$ are the ten-dimensional
gamma matrices, while $\gamma^a $ are the four dimensional ones. The
ten-dimensional spinor $\l$ is decomposed in terms of four-dimensional
spinors as
\be
\lambda = \left(
\begin{array}{c}
    \lambda_+^A \\ 
    \lambda_{-A}\\ 
  \end{array}
\right),~~
\text{with}\quad  \lambda_+^A =  \left(
\begin{array}{c}
    \Psi_\a \\ 
    0\\ 
  \end{array}
\right),
\ee
where the $SU(4)$ index $A=1,\ldots,4$ and $\l^A_+$ is a positive
chirality, four-dimensional spinor, so that $\Psi_\a$ carries an $SU(2)$
index $\a=1,2$, i.e. it is a complex 2-spinor. The negative chirality counter-part is
given by  $\l_{-A} = C_4(\bar{\l}_{+A})^T$, where $C_4$ is the
four-dimensional charge conjugation operator, see \cite{Ishiki:2006yr}
for details.
The covariant derivatives $D_a = \nabla_a -i[A_a,\hspace{.1cm}]$,  where 
\bsp
&\nabla_aA_b = e^\mu_a(\partial_\mu A_b + \omega_{\mu b}^{\hspace{.3cm} c}A_c),\\
& \nabla _aX^m = e^\mu_a\partial_\mu X^m,\\
&\nabla_a \lambda = e^\mu_a(\partial_\mu \lambda  
+ \frac{1}{4}\omega_\mu^{\hspace{.1cm}bc}\Gamma_{bc}\lambda),
\end{split}
\ee where $\mu = t,\theta, \phi, \psi$ is a curved-space index and
$a,b,c$ are tangent-space indices.  The non-vanishing components of the
vierbeins and spin connections are \bsp & e^1_\theta = 1/\mu,
\quad e^2_\phi = \frac{\sin \theta}{\mu}, \quad
e^3_\phi = \frac{\cos \theta}{\mu},
\quad e^3_\psi = 1/\mu,\\
& e^\theta_1 = \mu, \quad e^\phi_2 = \frac{\mu}{\sin \theta},
\quad e^\psi_2 = -\mu \frac{\cos \theta}{\sin \theta},
\quad e^\psi_3 = \mu,\\
&\omega_{12} = -\omega_{21} = -\frac{1}{2}(\cos \theta d\phi - d\psi),
\quad \omega_{23} = -\omega_{32} = -\frac{1}{2}d\theta,\\
&\omega_{31} = -\omega_{13} = - \frac{1}{2}\sin \theta
d\phi.
\end{split}
\ee
The supercharges $Q$, like the fermionic fields, are decomposed as   
\be Q = \left(
\begin{array}{c}
    Q_+^A \\ 
    Q_{-A}\\ 
  \end{array}
\right), \ee with $Q_{-A} = C_4(\bar{Q}_{+A})^T$.  Although the
$SO(6)$ basis, in which the scalar fields and gamma matrices are
represented as $X^m$ and $\G^m$, provides a compact expression for the
action (\ref{actions3}), we will need to express the supercharges of
the theory in an $SU(4)$ basis where we have instead $X^{AB}$,
$A,B=1,\ldots,4$, and similarly for the $\G^m$. The
dictionary between the two bases is given in appendix
\ref{sec:so6su4}.  We adopt a Hamiltonian formalism with $A_0 =0$.  In
the canonical formalism, the explicit expressions for the supercharges
are 
\bsp Q^*_{+A} = \frac{1}{g^2}\Tr\int_{S^3}\Biggl[ &g^2 \lambda
^*_{+A}\gamma^i E_i + \frac{1}{2} \lambda
^*_{+A}\gamma^{ij}\gamma^0F_{ij} -2 g^2\Pi_{AB}\lambda^{*B}_-\gamma^5-
2
(D_iX_{AC})\l^{*C}_-\gamma^i\gamma^5\gamma^0 \\
& \pm 2i\left(\frac{\mu}{2}\right)X_{AC}\l^{*C}_-\gamma^5
-2i[X_{AL},X^{LP}]\l^{*}_{+P}\gamma^0\Biggr],
\end{split}
\ee
\bsp\label{QpN4}
 Q_+^A = \frac{1}{g^2}\Tr\int_{S^3}\Biggl[&g^2\gamma^iE_i\l_+^A +
 \frac{1}{2} \gamma^0\gamma^{ij}\l^A_+F_{ij}  -2g^2\Pi^{AB}\g^5\l_{-B}
 + 2(D_iX^{AC})\g^0\g^5\g^i\l_{-C} \\
 & \mp 2i\left(\frac{\mu}{2}\right)X^{AM}\g^5\l_{-M} + 2i[X^{AL},X_{LP}]\g^0\l_+^P\Biggr],
\end{split}
\ee
where $E_i = g^{-2}\dot A_i$, $\Pi^{AB} =
g^{-2}\dot{X}^{AB}$, and where we have introduced the spatial
index $i,j=1,\ldots,3$ so that $a=(0,i)$.

The supersymmetry variation of a generic field $\mathcal{W}
\rightarrow \delta_\epsilon \mathcal{W} = [\bar{Q}_{+A}\epsilon^A_+ +
\bar{Q}_-^A\epsilon_{-A}, \mathcal{W}]$, where the spinor $\epsilon$
satisfies the conformal Killing equation  
\be \nabla_\mu \epsilon^A_+
= \pm \frac{i\mu}{4}\gamma_\mu \gamma^0\epsilon^A_+ .
\ee The two signs
on the $r.h.s.$ of the Killing equation result in the signs in front of
the ``mass-terms'' (the terms linear in $X_{AB}$ which do not involve
derivatives) in the expressions for the supercharges presented above.
In what is to follow, we shall take the upper sign in the Killing
equation, which would correspond to the lower sign in front of the
``mass-terms'' in the expression for the supercharges.

The canonical commutation relations following from the action are given by
\bsp\label{comrels}
&[X_{AB}(x), \dot X^{CD}(y)] = ig^2\frac{1}{2}\d^3(x-y)\, 
\left(\d^C_A \d^D_B - \d^C_B \d^D_A \right),
\\
&[A_i(x), \dot{A}_j(y)] = ig^2\d^3(x-y)\,\d_{ij},\\
&\{ \l_{+A}(x), \l^{\dag B}_{+}(y) \} = g^2\d^3(x-y)\,\d^{B}_A,\\
&\{ \l_{-}^A(x), \l^{\dag}_{B-}(y) \} = g^2\d^3(x-y)\,\d^{A}_B.
\end{split}
\ee
It is useful to extract the action of the supercharges on
single-particle states formed by the scalar and fermionic partons of
the theory.  For this purpose it is useful to introduce the
oscillators \bsp &\alpha^{AB} = \sqrt{\frac{\mu}{2g^2}}X^{AB} +
i\frac{1}{\sqrt{\frac{\mu}{2g^2}}}\Pi^{AB}, \hspace{.3cm}
\alpha^{\dagger}_{AB} = \sqrt{\frac{\mu}{2g^2}}X_{AB} -
i\frac{1}{\sqrt{\frac{\mu}{2g^2}}}\Pi_{AB},\\
&[\alpha^{AB},\alpha ^\dagger_{CD}] = \d^3(x-y)\left(\delta^A_C\delta^B_D -
\delta^A_D\delta^B_C\right).
\end{split}
\ee Notice, that these oscillators differ from the oscillator
variables usually employed in the canonical quantization of massive
scalar fields. We have $not$ Fourier decomposed any of the spacetime
coordinates, and the oscillator variables depend on the three $S^3$
coordinates as well as on time. The vacuum of the field theory is
taken to be annihilated by $\alpha_{AB}$ and $\lambda_+$.
On the single particle  states built out of the scalar and fermionic fields
\bsp
&[Q^A_+, \alpha^\dagger_{MN}]|0\rangle  =  2\sqrt{\frac{\mu}{2g^2}}\left(+ \delta ^A_N \left(
\begin{array}{c}
    0 \\ 
    \sigma^2\Psi^*_M\\ 
  \end{array}
\right) - \delta ^A_M\left(
\begin{array}{c}
    0 \\ 
    \sigma^2\Psi^*_N\\ 
  \end{array}
\right)\right)|0\rangle,\\
&[Q^*_{+A}, \alpha^\dagger_{MN}] |0\rangle= 0,\\
&\{Q^A_{+\alpha}, \lambda^*_{B\beta}\}|0\rangle  =
\left[\left(g^2E_i\g^i_{\alpha \beta} +
    \frac{1}{2}F_{ij}(\g^0\g^{ij})_{\alpha \beta}\right)\delta^A_B +
  2i[X^{AL},X_{LB}]\g^0_{\alpha \beta}\right]|0\rangle, \\
&\{Q^*_{+A\alpha}, \lambda ^*_{+B\beta}\}|0\rangle =
-2i\sqrt{\frac{\mu g^2}{2}}(\g^5C_4\g^0)_{\alpha
  \beta}\alpha^\dagger_{AB}|0\rangle.
\end{split}
\ee The above relations are true only modulo the equations of motion
and spatial translations, as they would be for any supersymmetric
Yang-Mills theory.

To proceed further, it is instructive to fix our conventions such that
\be
\g^0 = -i \begin{pmatrix}
0  &I\\
I  &0
\end{pmatrix}, \hspace{.2cm} \g^i =  \begin{pmatrix}
0  &i\sigma^i\\
-i\sigma^i  &0
\end{pmatrix}, \hspace{.2cm} C_4 =  \begin{pmatrix}
-\sigma^2  &0\\
0  &+\sigma^2
\end{pmatrix}, \hspace{.2cm} \g^5 =  \begin{pmatrix}
I  &0\\
0  &-I
\end{pmatrix}.
\ee
The two bosonic and two fermionic states transforming  under
$SU(2)_R$ and $SU(2)_L$ can be taken to be
\bsp
|\phi_a\rangle = \a^\dagger_{4a(=1,2)}|0\rangle, \hspace{.3cm} 
|\psi_\alpha\rangle  = \Psi^*_{4\alpha}|0\rangle.
\end{split}
\ee A note about the positions of the fermionic $SU(2)$ indices is in
order.  $\Psi^M = \Psi^M_{\alpha}$ and $\Psi^*_M = \Psi^{*\alpha}_M$
are the natural positions of the $SU(2)$ index ``$\alpha $'' on the two
component complex spinor $\Psi$. However in creating the state
$|\psi_\alpha\rangle $ the index is lowered using $\Psi^*_\alpha =
\epsilon_{\alpha \beta}\Psi^{*\alpha}$.  It is also understood that
$\epsilon_{12} = -\epsilon^{12} =1$.  After restricting $A,B = 1,2$ on
$Q, Q^*$ and renaming the restricted supercharges $q^a_\alpha,
q^{*\alpha}_a$, we obtain the fundamental representation of $SU(2|2)$
which can be expressed manifestly as \bsp\label{su22-main}
&q^a_\alpha |\phi_b\rangle = -2i\sqrt{\frac{\mu}{2g^2}}\delta^a_b|\psi_\alpha\rangle,\\
&q^{*\alpha}_a |\phi_b\rangle  = 0,\\
&q^a_\alpha |\psi_\beta \rangle = 2\epsilon_{\alpha \beta}\epsilon^{ab}|[\phi_b,Z]\rangle,\\
&q^{*\alpha}_a|\psi_\beta\rangle = +2i\sqrt{\frac{\mu
    g^2}{2}}\delta^\alpha _\beta |\phi_a\rangle.
\end{split}
\ee
The canonical anti-commutation relation between the supercharges is given by
\be\label{mass-algebra}
\{q^{a}_{\alpha},q^{*\beta}_{b}\} = 2\delta^a_b\delta_\alpha ^\beta H
+ 4\left(\frac{\mu}{2}\right)\delta^\beta _\alpha \mathcal{R}^a_b + 4
\left(\frac{\mu}{2}\right)\delta^a_b\mathcal{L}_\alpha^\beta,
\ee
where
\bsp
&\mathcal{R}^a_b = \sum_{C=1}^4\alpha^\dagger_{bC}\alpha^{aC} -
\delta^a_b\frac{1}{4}\sum_{M,N=1}^4\alpha^\dagger_{MN}\alpha^{MN},\\
& \mathcal{L}^\alpha_\beta = \Psi^{*\alpha}_4\Psi^4_\beta.
\end{split}
\ee
This completes our discussion of a concrete realization of
(\ref{radialsusy}) in the context of the Hamiltonian formulation of
$\mathcal{N}=4$ SYM on $\mathbb{R}\times S^3$.  As expected, the
r\^ole of the conformal dimensions is assumed by the masses  of the
collective excitations of the gauge theory in the curved background,
with the scale of the masses  being set by $\mu$.  

\subsection{Dimensional reductions}

We shall now work out the dimensional reductions of the Hamiltonian
and the supercharges to $\mathbb{R}\times S^2$. To carry out the
dimensional reduction to the three dimensional spacetime, we need to
assume that the scalar and fermionic fields do not depend on the
$U(1)$ coordinate $\psi$. As for the gauge field, one simply replaces
the third component of the one-form on the tangent space by a scalar;
namely $A_3 = \Phi$. More concretely \bsp
A &= A_ae^a_\mu dx^\mu = e^\mu_aA_\mu e^a_\nu dx^\nu\\
&= A_tdt + A_\theta d\theta + A_\phi d\phi + \frac{\Phi}{\mu}(\cos
\theta d\phi + d\psi).
\end{split}
\ee
Using this decomposition in (\ref{actions3}) and dropping the $\psi$
dependence of all the fields, we get
\bsp
S = \frac{1}{g^2_{S^2}} \Tr \int_{\mathbb{R}\times S^2} \Biggl[ &-\frac{1}{4} F_{ab}^2  -
\frac{1}{2} D_a \Phi D^a \Phi - \frac{\mu^2}{2}\Phi^2 + \mu F_{12}\Phi\\
&-\frac{1}{2} D_a X_{m} D^a X^m
-\frac{1}{2}\left(\frac{\mu}{2}\right)^2 X_mX^m -\frac{ i}{2} \bar \l
\Gamma^a D_a \l\\
&+ \frac{i\mu}{8}\bar{\lambda}\Gamma^{123}\lambda  -
\frac{1}{2}\bar{\lambda}\Gamma ^3[\Phi,\lambda]\\
&+\frac{1}{2}[\Phi,X_m]^2+\frac{1}{4}[X_m,X_n]^2
-\frac{1}{2}\bar \l \Gamma^m[X_m,\l] \Biggr].
\end{split}\label{actions2}
\ee The radius of $S^2$ is $1/\mu$, with the non-vanishing
dreibeins and spin connections given by the standard formulae \bsp
&b^1_\theta = \frac{1}{b^\theta_1} = \frac{1}{\mu},\hspace{.2cm}
b^2_\phi = \frac{1}{b^\phi_2} = \frac{\sin \theta}{\mu},\\
&\omega_{12} = -\omega_{21} = -\cos \theta d\phi.
\end{split}
\ee The three dimensional coupling $g^2_{S^2}$ is related to the four
dimensional one as \be g^2_{S^2} = \frac{g^2 \mu}{4\pi}.  \ee
The measure of integration
\be
  \int_{\mathbb{R}\times S^2} \equiv
\frac{1}{\mu^2} \int dt\int _0^\pi \sin \theta d \theta \int_0^{2\pi}
d\phi,
\ee
is defined to yield a volume of $4\pi/\mu ^2$.  This
dimensional reduction yields the following expression for the
supercharge for the three dimensional theory 
\bsp\label{QpN2} Q_+^A =
\frac{1}{g^2_{S^2}}\Tr\int_{S^2}\Biggl[&g^2_{S^2}\gamma^iE_i\l_+^A +
\frac{1}{2} \gamma^0\gamma^{ij}\l^A_+F_{ij}
-2g^2_{S^2}\Pi^{AB}\g^5\l_{-B}\\& + 2(D_iX^{AC})\g^0\g^5\g^i\l_{-C} 
 + 2i\left(\frac{\mu}{2}\right)X^{AM}\g^5\l_{-M} \\
&+2i[X^{AL},X_{LP}]\g^0\l_+^P + g^2_{S^2}\Pi_{\Phi}\g^3\lambda^A_+\\&
+\g^0\g^{i3}D_i\Phi\lambda^A_+ - i\mu\Phi \g^3\lambda^A_+
-2i[\Phi,X^{AC}]\g^0 \g^5\g^3\lambda_{-C}\Biggr],
\end{split}
\ee 
The last line contains all the terms involving the new (seventh)
scalar field obtained from the dimensional reduction of the four
dimensional vector potential.

Comparing the expression for the dimensionally reduced supercharge
with (\ref{QpN4}), it is easy to see that it admits a restriction to
an $SU(2|3)$ sector, just like (\ref{su22-main}). However, we have not
yet shown that the supercharge presented above is indeed a symmetry of
the Hamiltonian obtained by the dimensional reduction to
$\mathbb{R}\times S^2$.  To do that, we need to reproduce the
supercharge as the time component of a supercurrent, which we shall
now proceed to do.  To this end, it is instructive to recall that for
${\cal N}=1$ SYM in 10-d,
\be
{\cal L} = -\frac{1}{2} F_{MN} F^{MN} + i \bar \Psi \G^M D_M \Psi,
\ee 
the SUSY variations of the fields 
\be
\d A_N = -2i\bar\Psi \G_N \e,\qquad
\d \Psi = F_{PQ} \G^{PQ} \e,
\ee
produce the supercurrent
\bsp
j^M  &=2i\bar\Psi\G^M\G^{PQ} F_{PQ}\e.
\end{split}
\ee
The supercharge is then given by
\be
Q = \int_{\text{space}} j^0 = 2i  \int_{\text{space}} \bar\Psi \G^0\G^{PQ}F_{PQ}\e.
\ee

Keeping the ten dimensional theory in mind, one can write the Lagrangian for the $\bR\times S^2$
theory in the following form \cite{Grignani:2007xz}
\be
{\cal L} = -\frac{1}{2} F_{MN} F^{MN} + i \bar \Psi \G^M \nabla_M \Psi
-i\frac{\m}{4}\bar\Psi \G^{123} \Psi + 2\m\Phi F_{12} -
\frac{\m^2}{4} \phi_{\bar m}^2 -\m^2 \Phi^2,
\ee
where the 10-d gauge field is understood as $A_M =(A_\m,\phi_{m})=
(A_\m,\Phi,\phi_{\bar m})$ with $\m=0,1,2$ and $m=3,\ldots,9$,
$\bar m =4,\ldots,9$.  The SUSY variations of the
fields are expressed as
\be
\d A_M = -2i\bar\Psi \G_M \e,\qquad
\d\Psi = F_{MN}\G^{MN} \e - \m \G^m \G^{123}\phi_m \e - \m
\G^3\G^{123} \Phi \e.
\ee
Since the kinetic part of the action is not different from ${\cal
  N}=1$ SYM in 10-d the $\m$-independent part of the supercurrent that
does not involve total derivatives will be the same as in the ten
dimensional theory.  Extra total derivative (surface terms) are
generated from the $\m$-dependent piece of the variation of the
fermion kinetic term, whose contribution to $\d{\cal L}$ is
\be
\nabla_M\left(i\bar\Psi\G^M\left(\m \G^m \G^{123}\phi_m  + \m
\G^3\G^{123} \Phi \right)\e\right).
\ee
Of course the very same term is generated by the new $\m$-dependent
piece of $\frac{\d {\cal L}}{\d(\p_M\Psi)} \d \Psi$. These add in the
expression  for the supercurrent, giving us
\be
Q = \int_{S^2} \Biggl( 2i\bar\Psi \G^0\G^{PQ}F_{PQ} \e
-2i\m\bar\Psi\G^0\G^m\G^{123}\phi_m\e
-2i\m\bar\Psi\G^0\G^3\G^{123}\Phi \e \Biggr),
\ee
which is thus the supercharge for the $\bR\times S^2$ theory, albeit
in a rather compact notation.  Expressing the $SO(6)$ fields in terms
of $SU(4)$ ones using the dictionary in appendix \ref{sec:so6su4}, we find
\bsp
Q_+^A = &4i\int_{S^2}\Biggl( 
\frac{1}{2}\g^{\m\n}\g^0F_{\m\n}\,\l_+^A 
-2\g^\m\g^0 D_\m X^{AB} \l_{-B}
+\g^\m\g^3\g^0 D_\m \Phi \,\l_+^A\\
&+2ig\g^3\g^0[\Phi,X^{AB}]\l_{-B}
+2ig\g^0[X^{AC},X_{CB}]\l_+^B
-i\m X^{AB}\l_{-B}
-i\m \g^3 \Phi \,\l_+^A\Biggr).
\end{split}
\ee
This agrees with (\ref{QpN2}) up to the overall $4i$ outside, which
can be easily absorbed in a redefinition of the charge.

The construction clearly shows that we can restrict the three
dimensional theory consistently to an $SU(2|3)$ sector.
Furthermore, the reduced supercharges act on the $SU(2|3)$ states
exactly as in (\ref{su22-main}), with $g$ replaced by $g_{S^2}$. The
three dimensional supercharges constructed above satisfy the same
massive algebra (\ref{mass-algebra}) as the four dimensional theory
allowing us to constrain its quantum spectrum on algebraic grounds as
discussed below.

\section{Dispersion relations and the extended $SU(2|2)$ algebra}\label{su22-gauge}
In this section we shall focus on constraining the spectrum of the
four and three dimensional sixteen supercharge theories in the scheme
of radial quantization. We shall put special emphasis on the r\^ole
played by the scale introduced by the radius of the sphere
$1/\mu$. Following that we shall extend the formalism to
incorporate the three dimensional $\mathcal{N}=8$ theory.  Following
\cite{Beisert:2005tm, Beisert:2006qh} we write the $SU(2|2)$ algebra
(\ref{su22-main}) abstractly as \bsp\label{su22-abstract} &
q^a_\alpha|\phi_b\rangle = a\,\delta
^a_b|\psi_\alpha\rangle,\\ &q^{*\alpha}_a |\phi_b\rangle =
c\,\epsilon_{ab}\epsilon^{\alpha \beta}|\psi_\beta \rangle
,\\ &q^a_\alpha |\psi_\beta \rangle = b\,\epsilon_{\alpha
  \beta}\epsilon^{ab}|\phi_b\rangle,\\ &q^{*\alpha}_a|\psi_\beta\rangle
= d\, \delta^\alpha _\beta |\phi_a\rangle.
\end{split}
\ee 
In the fundamental representation, which corresponds to the
tree-level field theory, one has $a = -2i\sqrt{\mu /2g^2}$, $b=c=0$,
and $d= +2i\sqrt{\mu g^2/2}$. To proceed beyond the classical theory,
one needs to augment this algebra by two new generators $P, K$ defined
by \cite{Beisert:2005tm, Beisert:2006qh} \bsp
& \{q^a_\alpha, q^{b}_\beta\} = \epsilon^{ab}\epsilon_{\alpha \beta}P =   \epsilon^{ab}\epsilon_{\alpha \beta} ab \\
& \{q^{*\alpha} _a, q^{*\beta}_b\} = \epsilon^{\alpha
  \beta}\epsilon_{ab}K =  \epsilon^{\alpha
  \beta}\epsilon_{ab} cd
\end{split}
\ee where $P$ and $K$ annihilate physical states. Notice that unlike
the case of the gauge theory on $\mathbb{R}^4$, $P$ and $K$ are not
independent, as $P = K^*$. This is yet another artifact of the effect
of the conformal transformations that map the theory from flat
spacetime to the sphere. The conformal transformation maps the
superconformal generators to the conjugates of the supercharges, and
the relation between $P$ and $K$ is another
reflection of the same map.

Closure of the algebra on $H$ and the rotation generators yields \be H
= \frac{1}{4}(ad +bc) \hspace{.3cm} \mbox{and} \hspace{.3cm} (ad - bc)
= 2\mu. \ee The second, level-shortening condition, is easily checked
to be satisfied at the classical level, using the values of
$a,b,c$, and $d$ in (\ref{su22-main}). These relations also yield the
dispersion relation for the magnons \be H = \frac{1}{2}\sqrt{\mu
  ^2 + PK}. \ee All the statements made above hold both for the four
and three dimensional gauge theories.  However, for the specific case
of the three dimensional theory, we would like to emphasize that its
Hamiltonian involves the ``perfect square'' term $ (F_{12} - \mu
\Phi)^2$, whose minima generate the moduli space of the vacua of this
theory.  Our results for SYM on $\mathbb{R}\times S^2$ only apply to
the trivial vacuum $\Phi = 0$. From the viewpoint of the gauge-gravity
duality, the string dual (\ref{LM}) for the theory proposed in \cite{Lin:2005nh}
and studied later in the paper applies only to this vacuum, making it
particularly interesting\footnote{Other vacua of the gauge theory
  correspond to monopole backgrounds \cite{Ishiki:2006yr}. It would
  doubtless be interesting to understand the quantum spectra of the
  theory around these non-trivial vacua as well.}.
\\

To proceed further, it is important to  parameterize the algebra as
\bsp
& a = \sqrt{\mu h(\lambda)}\eta\\
& b =  \sqrt{\mu h(\lambda)} \frac{\zeta}{\eta}(1- x^+/x^-)\\
&c = \sqrt{\mu h(\lambda)} \frac{i\eta}{\zeta x^+}\\
&d = \sqrt{\mu h(\lambda)}\frac{x^+}{i\eta}(1- x^-/x^+)
\end{split}
\ee The above parameterization, where $h$ is an arbitrary function of
the dimensionless 't Hooft coupling of the gauge theory is specific to
the case of the $\mathcal{N}=4$ theory on $\mathbb{R}\times S^3$. That
is so because the length dimensions of the four parameters are
all equal to $-1/2$ only in the four dimensional theory.  It is easily
seen that for the $\mathbb{R}\times S^2$ case, $a$ and $c$ are
dimensionless, while $b$ and $d$ have length dimension $-1$.  The
appropriate parameterization in that case is \bsp
& a = \sqrt{h(\lambda)}\eta,\\
& b =  \mu \sqrt{ h(\lambda)} \frac{\zeta}{\eta}(1- x^+/x^-),\\
&c = \sqrt{ h(\lambda)} \frac{i\eta}{\zeta x^+},\\
&d = \mu\sqrt{h(\lambda)}\frac{x^+}{i\eta}(1- x^-/x^+).
\end{split}
\ee
However, in both cases the shortening condition implies
\be
\frac{2i}{h(\lambda)} = x^+ + \frac{1}{x^+} - x^- - \frac{1}{x^-}.
\ee
Also, writing $x^+/x^- = e^{ik}$ allows us to write 
\bsp
& P = ab = \mu h(\lambda)(1-e^{ik}),\\
& K = cd = \mu h(\lambda)(1 - e^{-ik}),
\end{split}
\ee for both theories in question. Thus the dispersion relation
for both the $\mathcal{N}=4$ and $\mathcal{N}=8$ theories can
naturally be expressed as \be H = \frac{\mu}{2}\sqrt{1 +
  4h^2(\lambda)\sin ^2(k/2)}. \ee Obviously, in the case of the three
dimensional theory, the dimensionless 't Hooft coupling is
$g^2_{S^2}N/\mu$. Using this expression, we see that the
dispersion relation agrees with the $k\ll1$ limit of the one derived
in \cite{Lin:2005nh} (and reviewed in section \ref{sec:string}) from
the world-sheet point of view.

The function $h(\lambda)$ cannot be fixed be from the
constraints of supersymmetry alone. In the following chapters we
determine it for the weakly coupled three dimensional theory at two
loop order (up to a single undetermined constant) and at strong
coupling from the dual string picture. In the gauge theoretic
analysis, we use the known results about the spectrum of the
dilatation operator of the four dimensional theory as a benchmark for
calibrating our methods and results.

\section{Weak coupling spectrum and integrable spin chains}\label{weak}

The form of the dispersion relation, discussed in the previous section
could be determined from an understanding of the realization of the
$SU(2|2)$ algebra alone. It has the same ``universal'' form for all the
dimensional reductions of the four dimensional theory whose
Hamiltonians are embedded in the $SU(2|2)$ structure as in
(\ref{su22-main}). We shall now focus on the determination of the
specific effective Hamiltonians for the gauge theories in question,
and show how their spectra are related to those of quantum spin
chains.  The new results in this section include the determination of
$h(\lambda)$ to two loop order in the three dimensional gauge theory,
which also appears to be integrable (at least in the $SU(2)$ sector)
at this order. We also determine the full $SU(2|3)$ symmetric
effective Hamiltonian at one-loop and find its leading
correction\footnote{For the full $SU(2|3)$ sector, there is an
  intermediate `dynamical' contribution between the one and two loop
  contributions.}.

To compute the effective Hamiltonians for both the
$\mathbb{R}\times S^3$ and $\mathbb{R}\times S^2$  within the
scheme of radial quantization we first recall that 
the states (for both theories) under consideration are generically of the form \beq |i_1
i_2\cdots i_n\rangle =
\frac{1}{\sqrt{N^{n}}}\Tr(a^\dagger_{i_1}a^\dagger_{i_2}\cdots
a^\dagger_{i_n})|0\rangle, \qquad a^\dagger_i =
(\alpha^\dagger_{4i})_0.  \eeq
$(\alpha^\dagger_{4i})_0$ corresponds to the lowest spherical harmonic
mode in the momentum space expansion of the oscillators
$(\alpha^\dagger_{4i})$, $i=1,2,3$ on $S^3$ or $S^2$. Though not
displayed above, it is implied that we also include the fermionic
states $\psi_\alpha$ so that the Hilbert space transforms under
$SU(2|3)$.

As it stands, although these states have a global $SU(N)$ invariance,
they do not seem to be invariant under local gauge transformations,
which will typically mix the different momentum modes. However, we
need to keep in mind that we shall work with a gauge fixed
Hamiltonian, for which the states can only be classified by their
quantum numbers. In such a gauge fixed $J^{PC}$-like scheme these
states are physical and normalizable. In the conformal field theory,
these states are mapped to local composite operators built out of
scalar fields alone once the theory is mapped to $\mathbb{R}^4$. Local
operators with covariant derivatives inserted on $\mathbb{R}^4$ would,
in turn, correspond to operators with higher spherical harmonics on
$\mathbb{R}\times S^3$.  The classification scheme for operators based
on $R$ charge and $J^{PC}$ assignments is valid for the three
dimensional $\mathcal{N}=8$ theory as well, as is the physicality of
the states mentioned above.

At tree level, the Hamiltonians for $\mathcal{N}=4$ SYM on
$\mathbb{R}\times S^3$ and $\mathbb{R}\times S^2$ reduce to harmonic
oscillator Hamiltonians, with a single oscillator assigned to each
angular momentum mode. The spectrum of the $SU(2|3)$ states is simply
given by their engineering dimensions at that level. The one-loop
correction to the energies is given by \be \Delta E^1 = \langle I|
:\left(H^4 + H^3\frac{\Pi}{E_0-H_0}H^3\right):|I\rangle = \langle
I|\Delta H^1|I \rangle ,\ee where $H^4$ and $H^3$ are the quartic and
cubic parts of the Hamiltonian.  $\Pi$ is the projector on to the
subspace\footnote{Note that this subspace includes states built
  from non-zero-mode excitations.} orthogonal to the states of energy $E_0$. The expressions for
$H^4, H^3$ are also taken to be normal ordered. The normal ordered
expressions can be mapped to Hamiltonians of quantum spin chains
\cite{Lee:1997dd, Lee:1998ea}. The general connection between matrix
models and their generalization to field theories and quantum spin
chains due to Lee and Rajeev has been reviewed in \cite{Lee:1999vb}.
For previous use of this identification in context of both the four
dimensional gauge theory and the plane wave matrix model we shall
refer to \cite{Agarwal:2004sz,Kim:2003rza, Klose:2003qc,
  Fischbacher:2004iu}. Here we recollect some of the relevant facts
about these matrix valued operators for the sake of
completeness.

The typical term at a given order in perturbation theory takes on the
form \be \Theta^I_J = \frac{1}{N^{(i+j-2)/2}}\Tr\left(W^{\dagger
    I_1}W^{\dagger I_{2}}\cdots W^{\dagger I_i}W_{J_j}
  W_{J_{j-1}}\cdots W_{J_1}\right). \ee The strings $I = I_1 I_2 \cdots
I_i$ and $J = J_1 J_2\cdots J_j$ denote fixed orderings of the bits
$I_i, J_j$, etc., which are shorthand for all $SU(2|3)$ and angular
momentum indices.  The oscillators $W_I$ collectively denote the three
bosonic and two fermionic matrix valued oscillator variables. These
$SU(N)$ invariant operators form a closed lie super-algebra, whose
basic anti-commutation relations are given as
\begin{eqnarray}
[\Theta ^I_J, \Theta ^K_L]_\pm &=& \delta ^K_J \Theta ^I_L  + \sum_{J
  = J_1J_2}(-1)^{\epsilon (J_1)[\epsilon (K) + \epsilon (L)]}\delta
^K_{J_2}\Theta ^I_{J_1L}\nonumber \\
& &+ \sum_{K = K_1K_2}\delta ^{K_1}_{J} \Theta ^{IK_2}_{L} +
\sum_{\stackrel{J=J_1J_2}{K=K_1K_2}}(-1)^{\epsilon (J_1)[\epsilon (K)
  + \epsilon (L)]}\delta ^{K_1}_{J_2}\Theta ^{IK_2}_{J_1L}\nonumber \\
& &+ \sum_{J = J_1J_2}\delta ^K_{J_1}\Theta ^I_{LJ_2} + \sum
_{J=K_1K_2}(-1)^{\epsilon (K_1)[\epsilon (I) + \epsilon (J)]}\delta
^{K_2}_J \Theta ^{K_1I}_{L}\nonumber \\
& & + \sum_{\stackrel{J=J_1J_2}{K=K_1K_2}}(-1)^{\epsilon
  (K_1)[\epsilon (I) +\epsilon (J)]}\delta ^{K_2}_{J_1}\Theta
^{K_1I}_{LJ_2}  \nonumber \\
& &+\sum_{J=J_1J_2J_3}(-1)^{\epsilon (J_1)[\epsilon (K) + \epsilon
  (L)]}\delta ^K_{J_2}\Theta ^I_{J_1LJ_3} \nonumber \\
& & + \sum_{K=K_1K_2K_3}(-1)^{\epsilon (K_1)[\epsilon (I) + \epsilon
  (J)]}\delta ^{K_2}_{J}\Theta ^{K_1IK_3}_{L}\nonumber \\
& & - (-1)^{[\epsilon (I)+ \epsilon (J)][\epsilon (K) +\epsilon (L)]}
\left(I,J \leftrightarrow K,L\right) .
\end{eqnarray}
In the above formula, expressions such as $\sum_{I = I_1I_2}$ imply
summing over all ways of writing the string $I$ as the concatenation
of two strings $I_1$ and $I_2$. $\epsilon(I)$ denotes the grade of the
string $I$, which is zero if it is bosonic and 1 if it is fermionic.
The full Lie algebra also includes an ideal which includes elements
that encode finite size effects. However, since we shall be working on
states of infinite size we shall ignore the contribution of the ideal,
which is irrelevant for our present concerns. A more complete
discussion of this algebra can be found in \cite{Lee:1999vb}.

When the operators are bosonic, their action on the single-trace
states can be expressed as 
\be \Theta ^I_J|K\rangle = \delta
^K_J|I\rangle + \sum _{K = K_1K_2}\delta ^{K_1}_J|IK_2\rangle .
\ee
Identifying the states with those of a quantum spin chain, we see that
$\Theta ^{ij}_{ji} = \sum_lP_{l,l+1}$, which is to be replaced by the
graded permutation operator $\Pi$ when fermionic creation and
annihilation operators are included.
 
On general grounds of $SU(2|3)$ invariance in the sector of the gauge
theories under consideration, the one loop effective Hamiltonian
$\Delta H^1 = \alpha \Theta^{ij}_{ji} + \beta\Theta^{ij}_{ij} =
\sum_l(\alpha \Pi_{l,l+1} + \beta I_{l,l+1})$, for some constants
$\alpha$ and $ \beta$. Requiring $\Delta H^1$ to annihilate the chiral
primary operators $\Tr(a^\dagger_3)^n|0\rangle$ yields $\alpha =
-\beta$.  To determine the coefficient of $\Pi$, we see that in the
bosonic $SU(2)$ sector the permutation operator arises entirely from
the quartic interaction vertex in $H^4$, whose contribution is \be
-\frac{1}{4g^2}\int_{\Omega} \Tr\left([X_{AB},X_{CD}][X^{AB},
  X^{CD}]\right) \rightarrow -\frac{g^2}{|\Omega
  |\mu^2}\Tr(a^\dagger_aa^\dagger_ba^aa^b) = -\frac{Ng^2}{|\Omega
  |\mu^2}\sum_l P_{l,l+1}. \ee The formula is true for both $S^2$ and
$S^3$, where $|\Omega |$ denotes the associated volume. Substituting
the explicit formulae for the volumes, we have \be \D H^1 = \frac{g^2N
  \mu}{16\pi^2}\sum_l(I - \Pi_{l,l+1}),\label{1loop} \ee for both the
theories in the closed $SU(2|3)$ sector. The coupling constant, for
the three dimensional theory expressed in terms of $g^2_{S^2}$ is
$g^2_{S^2}N/(4\pi)$. It is gratifying to note that for the four
dimensional theory, the above formula agrees with the known result for
the dilatation operator on $\mathbb{R}^4$ \cite{Minahan:2002ve}, after
one sets the radius of $S^3$ to unity,
i.e. $\mu =2$. It also agrees with the one loop result obtained for the three
dimensional theory in \cite{Ishiki:2006rt}. In that paper, $\Delta
H^1$ was computed for the full $SO(6)$ sector, which is closed, as it
is in the four dimensional theory, (only) at one loop. Restriction of
that result to the $SU(2)$ sector agrees with above Hamiltonian.
Moreover, our understanding of how the $SU(2|3)$ symmetry is realized
in the radial Hamiltonian formalism allows us to generalize the
one-loop result to the full $SU(2|3)$ sector and, as we shall see
below, go beyond the one-loop level.

\subsection{Higher loops}

In \cite{Beisert:2003ys}, it was shown how the $SU(2|3)$ symmetry
alone can be used to constrain the form of the higher loop corrections
to the dilatation operator of $\mathcal{N}=4$ SYM.  Although the four
dimensional superconformal theory was the focus of the analysis in
that paper, the results in \cite{Beisert:2003ys} can be readily
adapted to determine the leading corrections to the one-loop radial
Hamiltonians for the gauge theories we study as well. Requiring that
the generators of the $SU(2|3)$ algebra close order by order in
perturbation theory \cite{Beisert:2003ys} it is possible to write down
the complete leading correction to (\ref{1loop}) as \be \Delta
H_{SU(2|3)} = \frac{\mu}{2}\left(\lambda (\Theta ^{AB}_{AB} - \Theta
  ^{AB}_{BA}) - \sqrt{\frac{(\lambda)^3}{2}}(\epsilon^{\alpha
    \beta}\epsilon_{abc}\Theta ^{abc}_{\alpha \beta} +
  \epsilon_{\alpha \beta}\epsilon^{abc}\Theta _{abc}^{\alpha \beta}
  )\cdots \right) ,\ee where the capital indices in the first term on
the $r.h.s.$ are meant to stand for both the $SU(3)$ (i.e. $a$, $b$,
$c$, $\ldots$) and $SU(2)$ ($\a$, $\b$, $\ldots$) indices. The second
term breaks the $SU(2|3)$ invariance to $SU(2)\times SU(3)$ and it
encodes the ``dynamical'' effect of altering the length of the spin
chain. It is a non-trivial fact that the effective Hamiltonian is
integrable at this order \cite{Agarwal:2005jj}. In the context of the
four dimensional gauge theory, we have reproduced the known result for
the dilatation generator explicitly within the scheme of radial
quantization. However since both the form of $\Delta H_{SU(2|3)}$
given above and its integrability follows directly from the symmetry
algebra, we also claim the above formula to be the complete first
non-trivial correction to the effective Hamiltonian for the
$\mathcal{N}=8$ model on $\mathbb{R}\times S^2$. Furthermore, it also
appears to be integrable at this order.

The coupling constant $\lambda$ is to be identified with
$g^2N/(8\pi^2)$ for the four dimensional theory and
$g^2_{S^2}N/(2\mu\pi)$ for the $\mathbb{R}\times S^2$ model.

To analyze the question of integrability at two loops it is
instructive to restrict ourselves to the bosonic $SU(2)$ sector.  For
the four dimensional $\mathcal{N} =4$ theory, the explicit forms of
the dilatation operator, are known up to five loop order
\cite{Beisert:2003tq, Beisert:2003ys, Beisert:2007hz,
  Fiamberti:2009jw}.  These results can be readily mapped to the five
loop effective radial Hamiltonian using the maps between the 't Hooft
couplings of the theories on $\mathbb{R}^4$ and $\mathbb{R}\times S^3$
given before.  In the absence of alternate explicit computations of
spectra, one can continue to use the symmetry to constrain the radial
$SU(2)$ Hamiltonian for the $\mathbb{R}\times S^2$ model at the two
loop order up to a single undetermined constant. Stated explicitly \be
\Delta H_{SU(2)} = \frac{\mu}{2} \left(\lambda (\Theta ^{ab}_{ab} -
\Theta ^{ab}_{ba}) + \lambda^2[(2\Theta ^{ab}_{ba} -
  \frac{1}{2}\Theta^{abc}_{cba} - \frac{3}{2}\Theta^{abc}_{abc}) +
  \alpha_1(\Theta ^{ab}_{ab} - \Theta ^{ab}_{ba})] + \cdots\right),
\ee where $\alpha_1$ is the new undetermined constant for the three
dimensional theory which is equal to zero in the four dimensional
case, by the requirement of BMN scaling, which is present in the
$\mathcal{N}=4$ theory at this loop order. However, it is not known if
there is any reason to expect such a scaling in the three dimensional
theory as well. It is known that for the PWMM (as for ${\cal N}=4$
SYM), BMN scaling is violated only at the four-loop order
\cite{Fischbacher:2004iu,Beisert:2006ez}.  Nevertheless, the
perturbative integrability of the effective Hamiltonian is ensured for
arbitrary values of $\alpha_1$. A higher charge $\mathcal{Q} = \lambda
\mathcal{Q}_0 + \lambda^2\mathcal{Q}_1$ can be constructed such that
$[\Delta H, \mathcal{Q}] = \mathcal{O}(\lambda ^4)$. The explicit form
of the higher charge is \bsp &\mathcal{Q}_0 = \Theta^{cab}_{abc} -
\Theta^{bca}_{abc},\\ & \mathcal{Q}_1 = \left( -6\mathcal{Q}_0 +
\Theta^{dacb}_{abcd} - \Theta ^{bdca}_{abcd} + \Theta^{dbac}_{abcd} -
\Theta^{cbda}_{abcd}\right),
\end{split}
\ee and it establishes the two-loop integrability of the $SU(2)$
sector of the three dimensional theory. As we discuss below, the
scattering matrix of the spin chain describing the $SU(2|3)$ sector of
the gauge theory is factorized, which allows us to interpret the the
two loop integrability in the $SU(2)$ sector as an important piece of
evidence in favor of integrability of the full $SU(2|3)$ sector at
this perturbative order\footnote{The two-loop Hamiltonian for the full
  $SU(2|3)$ sector, determined up to a few constants, by requiring the
  perturbative closure of the algebra is available in
  \cite{Beisert:2003ys}. Those results are obviously valid for the
  three dimensional theory as
  well.}.
Finally, we note that at this order, the scaling function is
determined to be \be\label{hlambda} h^2(\lambda) = 2\lambda +2\alpha_1\lambda^2 +
\cdots .\ee
\section{Integrability and scattering matrices}
\label{sec:intS}

In this section we comment on carrying over the insights gained from
the studies of the multi-particle $S$-matrix for the planar dilatation
operator of $\mathcal{N}=4$ SYM on $\mathbb{R}^4$ to the radial
Hamiltonians described in the previous sections.  Having interpreted
the effective planar Hamiltonians in the $SU(2|3)$ sector of the three
and four dimensional gauge theories as spin chains, we can proceed to
constrain the $S$-matrix of the spin chain using the symmetry algebra
by adapting Beisert's techniques in \cite{Beisert:2006qh,
  Beisert:2005tm}. We bear in mind that the ferromagnetic vacuum of
the spin chains involves states made out of $Z$'s alone, while the
excitations/magnons transform under the residual $SU(2|2)$ symmetry,
which is also the symmetry of the $S$-matrix. Since the details of the
determination of the $S$-matrix by the use of the $SU(2|2)$ algebra
have been expanded on at length in \cite{Beisert:2006qh, Beisert:2005tm},
we shall refer to Beisert's original papers for the technical details.

The generalization of the single particle (fundamental) representation
of the $SU(2|2)$ algebra (\ref{su22-abstract}) to multiple particles
involves the introduction of the $\mathcal{Z}^{\pm}$ (length changing) markers as
follows \bsp\label{su22-z}
& q^a_\alpha|\phi_b\rangle = a\,\delta ^a_b|\psi_\alpha\rangle,\\
&q^{*\alpha}_a |\phi_b\rangle  = c\,\epsilon_{ab}\epsilon^{\alpha
  \beta}|\psi_\beta \,\mathcal{Z}^-\rangle, \\
&q^a_\alpha |\psi_\beta \rangle = b\,\epsilon_{\alpha
  \beta}\epsilon^{ab}|\phi_b\,\mathcal{Z}^+\rangle,\\
&q^{*\alpha}_a|\psi_\beta\rangle = d\,\delta^\alpha _\beta
|\phi_a\rangle.
\end{split}
\ee Comparison with (\ref{su22-main}) immediately clarifies the r\^ole
of the markers as essentially bookkeeping devices for gauge
transformations.  The generators $P$ and $K$ act as \be P|W\rangle =
ab|W\mathcal{Z}^+\rangle, \hspace{.3cm} K|W\rangle = cd|W{\cal Z}^-\rangle
.\ee In the case of the scattering of two magnons, the two particle
$S$-matrix can be constrained up to ten undetermined functions of the
magnon momenta. The action of the $S$-matrix on two particle states
can be expressed in all generality as \bsp &
S_{12}|\phi_a^1\phi_b^2\rangle =
A_{12}|\phi^2_{\{a}\phi^1_{b\}}\rangle +
B_{12}|\phi^2_{[a}\phi^1_{b]}\rangle
+\frac{1}{2}C_{12}\epsilon_{ab}\epsilon^{\alpha \beta}|\psi ^2_\alpha
\psi^1_\beta \mathcal{Z}^-\rangle \nonumber\\
& S_{12} |\psi_\alpha \psi_\beta\rangle =
D_{12}|\psi^2_{\{\alpha}\psi^1_{\beta \}}\rangle +
E_{12}|\psi^2_{[\alpha}\psi^1_{\beta ]}\rangle
+\frac{1}{2}F_{12}\epsilon_{\alpha
  \beta}\epsilon^{ab}|\phi_a^2\phi_b^1\mathcal{Z}^+\rangle\nonumber\\
& S_{12}|\phi^1_a\psi^2_\beta\rangle = G_{12}|\psi^2_\beta
\phi^1_a\rangle + H_{12}|\phi^2_a\phi^1_\alpha \rangle \nonumber\\
&S_{12}|\psi^a_\alpha \phi^2_b\rangle = K_{12}|\psi ^2_\alpha \phi
^1_b\rangle + L_{12}|\phi ^2_b \psi^1_\alpha\rangle
\end{split}
\ee Requiring the two body scattering matrix to commute with the
supersymmetry generators uniquely determines the ten undetermined
functions in terms of a single function $S^0_{12}$. For example
\be\label{S012} A_{12} = S^0_{12}\frac{x^+_2 - x^-_1}{x^-_2 - x_1^+}.
\ee The expressions for all the other functions $B\cdots L$ in terms
of $S^0_{12}$ can be found in table-1 of \cite{Beisert:2005tm}.
Furthermore, the scattering matrix satisfies the Yang-Baxter algebra
\be S_{12}S_{13}S_{23} = S_{23}S_{13}S_{12}, \ee fulfilling the
necessary condition for the integrability of the $SU(2|3)$ symmetric
spin chain to all orders in perturbation theory\footnote{The
  Yang-Baxter algebra is also a consequence of the Yangian symmetry
  exhibited by the $S$-matrix \cite{Beisert:2007ds}.}.  The magnon
momenta are to be determined by the Bethe ansatz equations obeyed by
the scattering matrix \cite{Beisert:2006qh, Beisert:2005tm}. For an
$m$ magnon state, the total energy is given by the additive relation
\be\label{HYM} H = \sum_{i=1}^m H_i = \sum_{i=1}^m
\frac{\mu}{2}\sqrt{1 + 4h^2(\lambda)\sin ^2(k_i/2)}.  \ee 
The
factorizability of the $S$-matrix and its determination up to a single
function are both consequences of the fact that the underlying
symmetry is $SU(2|2)$ and that the fundamental excitations fall on
atypical representations. The tensor product of this representation
uniquely gives a $single$ new irreducible representation, allowing us
to constrain the scattering matrix up to a single function of the
magnon momenta. The formal (matrix) structure of the two particle
scattering matrix, is hence a direct consequence of the symmetry and
the details of the models that realize these symmetries lie hidden in
the parameter $\mu$, the scaling function $h(\lambda)$ and, relatedly,
$S^0_{12}$.

While the Yang-Baxter condition on the $S$-matrix is a necessary
condition for integrability, it is certainly not sufficient.  One
typically needs to augment this with the existence of additional
conserved charges to gain surer evidence of integrability.  In the
case of the three dimensional gauge theory we have presented this
evidence in the form of higher conserved charges up to the two loop
order.  At least in the $SU(2)$ sector, one can hope to reliably use
the asymptotic Bethe ansatz techniques to compute the spectrum of the
three dimensional gauge theory at this loop order. The Yang-Baxter
relation satisfied by the full $SU(2|2)$ S-matrix suggests that Bethe
ansatz techniques may be applicable to a larger sector of the gauge
theory at and beyond two loops. Clearly, probing the structure of the
spin chain describing its spectrum at higher loop orders (even in the
$SU(2)$ sector) and understanding its integrability properties remains
an exciting open problem.

\section{Comments on PWMM and $\mathcal{N}=4 $ SYM on $\mathbb{R}\times S^3/{\bZ_k}$}
\label{sec:extend}
The principles used in constraining the spectrum of the three
dimensional gauge theory used above can also be employed in the study
of the PWMM and $\mathcal{N}=4 $ SYM on $\mathbb{R}\times
S^3/{\bZ_k}$.  These theories have a rich moduli space of vacua.
However, as long as a well defined large-$N$ expansion can be
implemented, the spectra around each of those vacua can, in principle,
be constrained by exactly the same use of the $SU(2|2)$ algebra as
explained above.  The different vacua would simply correspond to
different $h(\lambda)$.  For example, the $h$ function for the PWMM
around its trivial vacuum has been computed at the leading order on
the string theory side in \cite{Lin:2005nh} and up to the four loop
level in perturbation theory in \cite{Fischbacher:2004iu}\footnote{For
  an exposition of the perturbative realization of $SU(2|2)$ in weaky
  coupled PWMM see \cite{Aniceto:2008ep}. }. A different large-$N$
limit for the matrix model can also be taken if it is expanded around
its so-called fuzzy-sphere vacuum \cite{Maldacena:2002rb}. This
expansion simply maps the matrix model to the three dimensional gauge
theory studied above, and the corresponding $h$ would be the one
computed up to two loops in the preceding section.  Thus, while the
different $h$ functions are determined dynamically in these different
theories, the r\^ole of the underlying $SU(2|2)$ and the consequent
dispersion relation (\ref{HYM}) appears to be generic to the
dimensional reductions of $\mathcal{N}=4$ SYM on $\mathbb{R}\times
S^3$ that preserve the $SU(2|3)$ symmetry. Furthermore, the ``matrix''
structure of the $SU(2|2)$ S-matrix would also be the same for this
class of theories, with the specifics of the models and choices of
vacua being encoded in the dynamically determined observable $A_{12}$.
As mentioned before, these universal properties are however crucially
contingent on there being a systematic large-$N$ limit for the study
of the spectrum of any of these given models around a particular
vacuum, which is assumed to be ``well separated'' in the sense
described below\footnote{The well-separatedness of the vacua and
  formulations of large-$N$ limits are not independent issues
  \cite{Lin:2005nh, Maldacena:2002rb}.}.

\section{$SU(2|2)$ from string theory}
\label{sec:string}

In this section we will discuss the emergence of the $SU(2|2)$ algebra
in the string theories dual to the family of gauge theories with 16
supercharges discussed previously. We will succeed in deriving the
explicit algebra in the plane-wave (BMN-like) limit, i.e. in the limit
that the magnon momenta $k \ll 1$. This amounts to analyzing type IIA
strings on the plane-wave, for which exact quantization has been
performed. The main point of the analysis is to show the emergence of
the central charges through a relaxation of the level-matching
condition. This was originally understood for the full $AdS_5\times
S^5$ superstring in \cite{Arutyunov:2006ak}; we will have to content
ourselves with the plane-wave limit, as the full geometries relevant
to our cases are complicated and have so far not admitted a solvable
sigma model. We find it useful to first review the bubbling geometries
method which was used to construct the full dual supergravity
solutions -- the so-called Lin-Maldacena solutions -- in which the
problem is reduced to a classical axisymmetric electrostatics problem
\cite{Lin:2005nh}.  The results presented here and in sections
\ref{sec:LCGPW} and \ref{sec:supch} are known from the literature. Our
contribution -- deriving the $SU(2|2)$ algebra -- is presented in
section \ref{sec:rest}.

The geometries have the bosonic symmetry $\bR\times SO(3)\times
SO(6)$; they contain (in addition to the temporal direction) an $S^2$
and an $S^5$ whose radii vary with the remaining two coordinates $\r$
and $\eta$.  The geometries may be thought of as arising from M2 and
M5 branes wrapping these contractile spheres. These geometries also
contain a $(\r,\eta)$-dependent dilaton and B-field, of which the
latter has its legs in the $S^2$. There are also one-form and
three-form Ramond-Ramond potentials $C_1$ and $C_3$, similarly
dependent only on $(\r,\eta)$, for which $C_1 \propto dt$ and $C_3
\propto dt\w d\O_2$, where $d\O_2$ denotes again the $S^2$. The
$(\r,\eta)$ plane is the scene of the axisymmetric electrostatics
problem ($\r$ is the radial, and $\eta$ the axial coordinate), whereby
the electric potential $V(\r,\eta)$ comes to inform the specific
dependence of the geometry on those coordinates. More specifically the
electric potential is that found in the presence of a configuration of
``critically'' charged conducting disks centered on the $\eta$-axis,
subject to a certain asymptotically defined external electric field.
The ``critical'' charge corresponds to the condition that the charge
density exactly vanishes at the edge of the disks, which in turn
implies that the electric field is non-infinite there. This condition
on the charge, and the asymptotic form of the external potential, are
determined by requiring that the corresponding supergravity solutions
are well-behaved and non-singular. The total charge per disk and the
distance between disks are proportional to the units of $*dC_3$ and
$dB_2$ flux on six- and three-cycles constructed from the $S^5$ and
$S^2$ and a $\r$ or, respectively, $\eta$ fiber. The emerging picture
is very attractive, with the solutions in one-to-one correspondence
with disk configurations.
 
The disk configurations corresponding to the dual of SYM on $\bR\times
S^2$ are the simplest: finite radii disks, a single disk corresponding
to the trivial vacuum where all scalar fields have zero VEV. Adding
more disks corresponds to the other vacua of the theory, see
\cite{Lin:2005nh} for a discussion. SYM on $\bR\times S^3/\bZ_k$
consists of a periodic extension of this configuration, extending to
$\pm \infty$ in $\eta$. Finally, the dual of the PWMM is obtained from
an infinite disk at $\eta=0$ giving the trivial vacuum, other vacua
being obtained through the addition of finite-radii disks at $\eta
>0$. The simplest plane-wave limit of these geometries is given by the
IIA plane-wave with $SO(3)\times SO(4)$ symmetry
\bsp\label{pwmu}
&ds^2 = -2dx^+dx^- -\left(\left(\frac{\b}{3}\right)^2 x^ix^i
+   \left(\frac{\b}{6}\right)^2 x^{i'}x^{i'} \right) (dx^+)^2
+ dx^i dx^i + dx^{i'} dx^{i'},\\
&F_{+123} = \b = -3 F_{+4},
\end{split}
\ee
where $i$ and $i'$ run from 1 to 4 and 5 to 8, respectively, and where
$\b$ is an arbitrary positive constant. This geometry is obtained by
expanding the Lin-Maldacena geometries around the region corresponding
to the edge of a single disk, ensuring any other disks are
well-separated from this region; this is what is meant by
``well-spaced'' vacua in footnote \ref{foot:cav}.

\begin{figure}
\begin{center}
\includegraphics*[bb=60 65 350 300, height=2.0in]{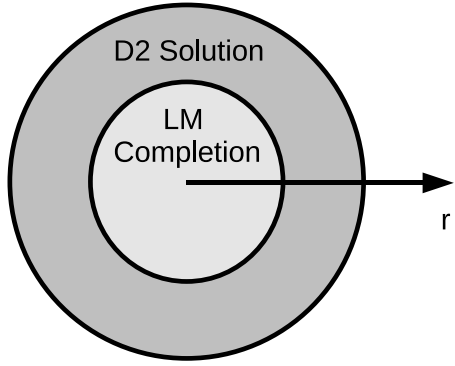}
\end{center}
\caption{The Lin-Maldacena solution dual to SYM on $\bR\times S^2$ is
  a completion of the D2-brane geometry \cite{Itzhaki:1998dd} to the
  IR region (small $r$).}
\label{fig:LMD2}
\end{figure}

Consistent with the focus of the paper, we will use SYM on $\bR \times
S^2$ around the trivial vacuum as the prototypical example, but of
course the analysis following (in sections \ref{sec:LCGPW} -
\ref{sec:stringgen}) is valid for the single-disk, plane-wave limit
dual of any of the gauge theories. The full supergravity solution dual
to the trivial vacuum of SYM on $\bR \times S^2$ is \cite{Lin:2005nh}
\bsp\label{LM}
&ds_{LM}^{2} = \a'L^{1/3} \Bigl[
 -8(1+r^{2})f dt^{2}+16{f}^{-1}\sin ^{2}\theta d\Omega _{5}^{2}\\
&\qquad+\frac{8r f }{
r+(1+r^{2})\arctan r}\left( \frac{dr^{2}}{1+r^{2}}+d\theta ^{2}\right) 
  +\frac{2r\left[r+(1+r^{2})\arctan r\right] f }{1+r\arctan r}d{\Omega }_{2}^{2}
\Bigr],\\
&f \equiv \sqrt{\frac{2}{r}[r+ (\cos ^{2}\theta +r^{2})\arctan
r]},\\
&B_{2} =- L^{1/3} \frac{2\sqrt{2}\left[ r+(-1+r^{2})\arctan r
\right] \cos \theta }{1+r\arctan r}d^{2}\Omega , \\
&e^{\Phi } = K L^{1/2}  8 r^{\frac{1}{2}}(1+r\arctan
r)^{-\frac{1}{2}} [ r+(1+r^{2})\arctan r]^{-\frac{1}{2}}f
^{-\frac{1}{2}} , \\
\end{split}
\ee
with 
\bsp
 &C_{1} =- K^{-1} L^{ -  \frac{1}{3} }
\frac{\left[ r+(1+r^{2})\arctan r\right] \cos \theta }{2r}dt,
\\
&C_{3} =-K^{-1}
\frac{r [ r+(1+r^{2})\arctan r] ^{2}f ^{2}}{\sqrt{2}(1+r \arctan r) }dt\wedge d^{2}\Omega,
\end{split}
\ee
where $L$ and $K$ are constants which will be related to gauge theory
parameters below. This solution may be viewed as an IR completion of
the D2-brane solution on $\bR\times S^2$ \cite{Itzhaki:1998dd}, which
suffers from a diverging dilaton as the radial coordinate $r$
approaches zero, invalidating the 10-dimensional description, see
figure \ref{fig:LMD2}. The D2-brane solution is given by
\be
ds_{D2}^2 = \a' C^{2/3} \left( r^{5/2} \left(-dt^2 +  d\O_2^2
\right) + \frac{dr^2}{r^{5/2}} + r^{-1/2} d\O_6^2 \right),\quad
e^{\Phi} = \frac{g^2}{\m C^{1/3}} \, r^{-5/4},
\ee 
where $C^2 = 6\pi^2 g^2 N/\m$, $g$ is the Yang-Mills coupling
constant, and $1/\m$ is the radius
of the gauge-theory $S^2$. Here we have used dimensionless coordinates
$t$ and $r$ in order to match-up with those of $ds^2_{LM}$. It will be
important for us later in matching to the gauge theory results that
$t=\m\, t_{YM}$, where $t_{YM}$ is the dimensionful gauge
theory time coordinate\footnote{The coordinate $r$ is related to the
  usual coordinate $U$ from \cite{Itzhaki:1998dd} by $U=C^{2/3}\m r$.}.  
Indeed taking $r\to\infty$ and scaling $r\to (8/\pi)^{1/3} r$, $t\to
t/2$ in $ds^2_{LM}$, $ds^2_{D2}$ is recovered with the following
identification of parameters
\be\label{gtquan}
L = \frac{3\pi^3\l}{2^9\sqrt{2}} ,\quad
K = \frac{4 g^2\sqrt{2}}{(6\l)^{2/3}\pi},\quad
t_{LM} = \frac{t_{D2}}{2} = \frac{ \m}{2} t_{YM},
\ee 
where we have introduced the dimensionless gauge theory 't Hooft
coupling $\l=g^2N/\m$. As was explained in \cite{Lin:2005nh}, this
identification gives $L$ and $K$ in terms of gauge theory quantities.
Plugging them into the Lin-Maldacena solution string coupling
$\exp(\Phi)$, one finds $\sim \l^{5/6}/N \times g(r,\theta)$,
where the function $g(r,\theta)$ is always finite (i.e. $\leq {\cal
  O}(1)$) and goes to zero for large-$r$ as $r^{-5/4}$. This implies
that at large-$N$, the string coupling is suppressed everywhere, and
so the solution may be trusted for any $r$.

The coordinate $r$ is related to the gauge theory energy scale. The
running of the SYM coupling constant is trivial and given simply by
dimensional analysis, so that the dimensionless effective coupling is
given by $g_{eff} = g^2N/E$, where $E$ is the relevant energy scale.
Since we have SYM on $\bR\times S^2$, it is more sensible to express
this scale in units of the $S^2$ radius $1/\m$, i.e.  ${\cal E} =
E/\m$, and so write $g_{eff} = \l/{\cal E}$. This running is reflected
in the string solution by the coordinate dependence of the string
coupling $\exp(\Phi)$.  This allows us to identify ${\cal E} \sim
g^{-6/5}(r,\theta)$ and so at large-$r$, ${\cal E}\sim r^{3/2}$. The
curvature scale of the geometry, for example the inverse radius of the
$S^5$, diverges for large-$r$. Thus the strongly curved region
corresponds to weak effective gauge coupling or large gauge theory
energies; we will therefore call this the UV region. In this part of
the geometry (for large-$N$), strings propagate classically on a
string-scale-curved background.  The small-$r$ or IR region
corresponds to strong effective gauge theory coupling and weak
curvature scales in the geometry, where a classical supergravity
analysis is appropriate.

If the large-$N$ limit is relaxed, the geometry may still be trusted
for large enough $r$, but at small $r$ the string coupling will be
large leading to a transition to an 11-dimensional description. For
small $N$, one expects then to make contact with the superconformal
M2-brane or Bagger-Lambert-Gustavsson
\cite{Bagger:2006sk,Bagger:2007jr,Bagger:2007vi,Gustavsson:2007vu}
theory (i.e. ABJM \cite{Aharony:2008ug} at $k=1,2$) and their massive
counterparts \cite{Hosomichi:2008jd,Hosomichi:2008jb,Gomis:2008cv,
  Gomis:2008vc}. Of course, we do not expect to find integrability at
finite $N$.

We are interested in taking a Penrose limit of the Lin-Maldacena
geometry around a stable light-like geodesic on the $S^5$. The
geodesic is a line in the $t$-$\phi$ plane, where $\phi$ describes a
great circle in the $S^5$.  This geodesic is located in the deep IR at
$r=0$, $\theta=\pi/2$, corresponding to the strongly coupled limit of
the gauge theory\footnote{This location in the $r$-$\theta$ plane
  corresponds to $|g_{tt}| = g_{\phi\phi}$. At this location we have a
  massless geodesic corresponding to $E=J$. Away from $r=0$,
  $\theta=\pi/2$, $E > J$ and the geodesic describes a particle of
  mass $m^2 = E^2-J^2$. Thus the chosen geodesic is a stable
  minimization of the energy \cite{Lin:2005nh}.}. The radius of the
$S^5$ in (\ref{LM}) at $r=0$ and $\theta=\pi/2$ is equal to
\be\label{Rdef}
\frac{R_{S^5}^2}{\a'} = 8  \sqrt{2} L^{1/3} =  \left( 6\pi^3\l\right)^{1/3}, 
\ee
where we have used (\ref{gtquan}) to express it in terms of gauge
theory quantities. In order to take the Penrose limit we define the
energy $E$ and angular momentum $J$ generators as $i\p_t \equiv E-J $
and $-i\p_\phi \equiv J$, and then define ($R^2=R^2_{S^5}$)
\bsp\label{xpmdef}
&x^+ = t,\quad x^-=R^2(t-\phi),\quad
p^- = i\p_{x^+} = i \p_t =E-J,\\
&p^+ = i\p_{x^-} = R^{-2}(i\p_t-i\p_\phi) = R^{-2} E \simeq R^{-2} J,
\end{split}
\ee
so that the light-cone energy is zero for $E=J$. We will take the
usual BMN-like limit by taking $R\to\infty$ and concentrating on
states with finite $p^+$ and $p^-$, so that $E \gtrsim J \sim R^2$.
Starting from (\ref{LM}) we take a Penrose limit around $r=0$, $\theta=\pi/2$
\bsp
&\theta = \frac{\pi}{2} + \frac{\sqrt{2}}{R} \, x^{i=1},\quad
r \Theta^{i=2,3,4}= \frac{\sqrt{2}}{R} \, x^{i=2,3,4},\\
&d\O_5^2 = \frac{dy^2}{(1+y^2/4)^2} + \frac{(1-y^2/4)^2}{(1+y^2/4)^2}
d\phi^2,\quad
 y^{i'} = \frac{x^{i'}}{R},
\end{split}
\ee
where $\vec\Theta$ is the embedding of the unit-$S^2$ which appears as
$d\O_2^2$ in (\ref{LM}) into $\bR^3$. This gives the IIA plane-wave
\be
ds^2 = -2dx^+dx^- -\left(4 x^ix^i
+  x^{i'}x^{i'} \right) (dx^+)^2 + dx^idx^i+ dx^{i'} dx^{i'} +{\cal O}(R^{-2}),
\ee
where one can match-up with (\ref{pwmu}) by setting\footnote{One may
  also verify that the Ramond-Ramond field strengths come out
  correctly. Note that the constant $\b$ may be absorbed into the
  coordinates and their relations to gauge theory parameters and has
  no physical significance in the gauge theory.} $\b = 6$. In the
following sections we will analyze the $x^{i'}$ excitations of strings
on this geometry. It is of course well-known that the energy
of such excitations are given by
\be
p^- = \sum_i  \sqrt{\left(\frac{\b}{6}\right)^2 + \frac{n_i^2}{(\a' p^+)^2}},
\ee
where $n_i$ are the worldsheet momenta of the excitations. Using
(\ref{gtquan}) and (\ref{xpmdef}), and setting $\b=6$ one obtains
\cite{Lin:2005nh}
\be
H_{YM} = \frac{\m}{2}\,p^-=\frac{\m}{2} \sum_i \sqrt{1 + \frac{R^4}{\a'^2} \frac{n_i^2}{J^2}},
\ee
which matches the gauge theory result (\ref{HYM}) in the $k_i =n_i/J \ll 1$
limit, and through (\ref{Rdef}), gives us the strong coupling limit of
the function $h(\l)$
\be
h(\l) \simeq \left(6\pi^3 \l \right)^{1/3}.
\ee

It is worth pointing out that while the three dimensional gauge theory only has sixteen  supersymmetries  the plane wave geometry is invariant under 24 supersymmetries. This is suggestive of an   enhancement of supersymmetry in the BMN limit of the strongly coupled gauge theory and a potential connection between the three dimensional strongly coupled SYM theory and $\mathcal{N}=6$ Chern-Simons models.

We will now continue with an analysis of the supersymmetry algebra of
strings on the IIA plane-wave, with the ultimate goal of uncovering
the $SU(2|2)$ structure found in the gauge theory. We begin with some
general known results. 

\subsection{Light-cone gauge strings on a plane-wave}
\label{sec:LCGPW}

The light-cone gauge quantization of strings on the plane-wave
geometry (\ref{pwmu}) corresponding to the Penrose limit described in the previous
section was carried out in the series of papers
\cite{Sugiyama:2002tf,Hyun:2002wp,Hyun:2002wu}. Using this analysis,
we will show that the $SU(2|2)$ algebra emerges from the commutation
relations of the supercharges. The main issue is to show that the
central charges emerge from a relaxation of the level-matching
condition, exactly as was shown for strings on $AdS_5\times S^5$ in
\cite{Arutyunov:2006ak}. We will find it useful to repeat portions of
the analysis in \cite{Hyun:2002wp}, both in the interest of
readability and because we will use slightly different conventions.

The string action is given by
\bsp
S = -\frac{1}{4\pi\a'} \int& d^2\s \,\Bigl\{ \sqrt{-h} h^{ab}\Bigl[
-2\p_a X^+ \p_b X^- + \p_a X^I \p_b X^I - M(X^I) \p_a X^+ \p_b X^+\\
&-2 \p_a X^+ \bar \T \G^- \p_b \T +  \p_a X^+ \p_b X^+
\Upsilon(\T) \Bigr]
+2 \e^{ab}\, \p_a X^+\, \bar \T \G^{-9} \p_b\T \Bigr\},
\end{split}
\ee
where the index $I=(i,i')$, where $i=1,\ldots,4$ and $i'=5,\ldots,8$,
and where we have introduced the shorthand
\be
M(X^I) = \left(\frac{\b}{3}\right)^2 X^iX^i 
+ \left(\frac{\b}{6}\right)^2 X^{i'}X^{i'} ,\qquad
\Upsilon(\T) = \frac{\b}{2}\,\bar\T \G^-\left(\G^{123} + \frac{1}{3}\G^{49}\right)\T.
\ee
We will hold off for the time being on the explicit details of the
fermions $\T$ and the 10-d gamma matrices $\G^A$. We continue by
calculating the Virasoro constraints, and then imposing the light-cone
gauge 
\be
h_{ab} = \text{diag}(-1,1),\quad X^+ = \a'p^+\t.
\ee
We find
\bsp\label{vircons}
&{X^-}' = \frac{1}{\a'p^+} \left( \dot X \cdot X' - \a' p^+ \bar\T \G^-
  \T' \right), \\
&\dot X^- = \frac{1}{2\a' p^+} \left( \dot X^2 + {X'}^2 - (\a' p^+)^2
  M(X^I) -2\a' p^+ \bar \T \G^-\dot \T + (\a' p^+)^2 \Upsilon(\T) \right),
\end{split}
\ee
where $\dot {} = \p_0 = \p_\t$ and $' = \p_1 = \p_\s$, and the inner
product is over the composite $I$ index. Using these expressions to
eliminate $X^-$ we may calculate the light-cone conjugate momenta. The
Lagrangian $L$ and Lagrangian density ${\cal L}$ are given as
\be
S = \int d\t \, L = \int d^2\s\, {\cal L},
\ee
from which we calculate
\bsp
&p^+ = \int_0^{2\pi} d\s\, P^+ 
= \int_0^{2\pi} d\s\,\left(-\frac{\d{\cal L}}{\d\dot X^-}\right),\\
&p^- = \int_0^{2\pi} d\s\, P^- 
= \int_0^{2\pi} d\s\,\left(-\frac{{\d\cal L}}{\d\dot X^+}\right)
\equiv H.
\end{split}
\ee
The first of these equalities is a tautology resulting from the
consistent definition of $X^+$, while the second yields the light-cone
Hamiltonian $H$, given by
\be
H= \frac{1}{4\pi{\a'}^2p^+} \int_0^{2\pi} d\s\, \Bigl(
 \dot X^2 + {X'}^2 + (\a' p^+)^2 M(X^I) 
+2\a' p^+ \bar \T \G^{-9} \T' - (\a' p^+)^2 \Upsilon(\T) \Bigr).
\ee

The first Virasoro constraint in (\ref{vircons}) yields one extra
piece of information, the level matching condition. It is this
condition which is relaxed in order to reveal the central charge of
the $SU(2|2)$ algebra. This issue has been worked out for the
$AdS_5\times S^5$ string in \cite{Arutyunov:2006ak}, where it has been
shown that the central charge, related as it is to changing the length
of the gauge theory spin-chain \cite{Beisert:2005tm}, appears in the
string treatment by going off-shell through a relaxation of the level
matching condition. Specifically, one takes the total worldsheet
momentum $p_{ws}$ to be
\be\label{levelmatch}
p_{ws} = \frac{1}{\a'p^+} \int_0^{2\pi}d\s
 \left( \dot X \cdot X' - \a' p^+ \bar\T \G^-
  \T' \right) \neq 0.
\ee
We will see that the supersymmetry algebra will allow us to associate
$p_{ws}$ with the gauge theory magnon momentum $k$.

\subsection{Supercharges and algebra}
\label{sec:supch}

The fermions $\theta$ are given by \cite{Hyun:2002wu}
\be
\T = \frac{1}{\sqrt{2\a' p^+}}\frac{1}{2^{1/4}} 
\begin{pmatrix} 0\\ \psi^A \end{pmatrix},
\ee
where the $\psi^A$ are 16-component real and are further decomposed
according to their $SO(8)$ and $SO(4)$ chiralities
\be
\g^9 \psi^1_{\pm} = + \psi^1_{\pm},\quad 
\g^9 \psi^2_{\pm} = - \psi^2_{\pm},\quad
\g^{1234} \psi^A_{\pm} = \pm \psi^A_{\pm}.
\ee
The gamma matrices and related conventions are collected in appendix
\ref{app:gammas}. There are 2 dynamical supercharges $Q_+^1$ and
$Q_-^2$ which have been constructed in \cite{Hyun:2002wp}, they are
\bsp
Q_+^1 = \frac{1}{4\pi\a'}\frac{1}{\sqrt{\a'p^+}}\int_0^{2\pi}d\s\,
\Bigl( &\p_-X^i \g^i \psi_-^1 + \frac{m}{3}X^i\g^i\g^4\psi_+^2 \\
&+ (i \to i',~ \psi_+ \lr \psi_-,~m \to -m/2) \Bigr),\\
Q_-^2 = \frac{1}{4\pi\a'}\frac{1}{\sqrt{\a'p^+}}\int_0^{2\pi}d\s\,
\Bigl( &\p_+X^i \g^i \psi_+^2 - \frac{m}{3}X^i\g^i\g^4\psi_-^1 \\
&+ (i \to i',~ \psi_+ \lr \psi_-,~m \to -m/2) \Bigr),
\end{split}
\ee
where $\p_\pm = \p_\t \pm \p_\s$, and $m=\b\a'p^+$. The canonical
commutation relations for the fields at equal times $\t$ are given by
\be
[X^I(\s), \dot X^J(\s')] = i 2\pi\a' \d^{IJ} \d(\s-\s'),\quad
\{\psi^A_\pm(\s), \psi^B_{\pm}(\s')\} = 2\pi \a' \d^{AB} \d(\s-\s').
\ee
In order to reveal the $SU(2|2)$ algebra we are interested in, we find
it necessary to define the following projected supercharges
\be
Q = \frac{e^{i\pi/4}}{\sqrt{2}}(1+\g^4)
\left( Q^1_+ + i Q^2_- \right),\qquad
\bar Q \equiv Q^* = \frac{e^{-i\pi/4}}{\sqrt{2}}(1+\g^4)
 \left( Q^1_+ - i Q^2_- \right),
\ee
and then to restrict $Q$ and $\bar Q$ to the
appropriate subalgebra. First let us quote the
result before the restriction. We find\footnote{We also find a
  contribution to $\{Q,Q\}$ and $\{\bar Q,\bar Q\}$ given by $ 
-\frac{i}{8\pi\a'} \frac{\b}{3}\int_0^{2\pi} d\s 
\left( \psi^1_-\psi^1_- -\psi^1_+\psi^1_+ + \psi^2_+\psi^2_+ -
\psi^2_-\psi^2_-\right)  (1+\g^4)_{\a\b}$, which is a vanishing sum of
$\d(0)$ infinities.}
\bsp\label{bigcomms}
&\{ Q_\a, \bar Q_\b\} = (1+\g^4)_{\a\b} \, H 
+ \frac{\b}{3} \, {\cal J}_{\a\g} (1+\g^4)_{\g\b} 
- \frac{\b}{6} \, {\cal J'}_{\a\g} (1+\g^4)_{\g\b}, \\
&\{ Q_\a, Q_\b \} = -i\frac{p_{ws}}{2\pi\a'}  (1+\g^4)_{\a\b},\\
&\{ \bar Q_\a, \bar Q_\b \} = i\frac{p_{ws}}{2\pi\a'}  (1+\g^4)_{\a\b},\\
\end{split}
\ee
where $\a,\b,\g$ are $SO(8)$ spinor indices ranging from 1 to 16, and where
\bsp
&{\cal J}_{\a\g} = \frac{i}{4\pi\a'} \int_0^{2\pi} d\s
\Bigl( \dot X^{\hat i} X^{\hat j} -  \dot X^{\hat j} X^{\hat i}\\
&\qquad\qquad\qquad
-\frac{i}{4} \left(  \psi^1_- \g^{\hat i \hat j} \psi^1_-
 +\psi^1_+\g^{\hat i \hat j}\psi^1_+ + \psi^2_+\g^{\hat i \hat j}\psi^2_+ +
\psi^2_-\g^{\hat i \hat j}\psi^2_-\right)\Bigr)\, \g^{\hat i
  \hat j}_{\a\g},\\
&{\cal J'}_{\a\g} = \frac{i}{4\pi\a'} \int_0^{2\pi} d\s
\Bigl( \dot X^{i'} X^{j'} -  \dot X^{j'} X^{i'}
-\frac{i}{4} \left(  \psi^1_- \g^{i' j'} \psi^1_-
 + \psi^2_+\g^{i' j'}\psi^2_+ +\right)\Bigr)\, \g^{i'
  j'}_{\a\g},\\
\end{split}
\ee
where we have introduced the $SO(3)$ index $\hat i = 1,2,3$. Notice
the crucial observation that the centrally extended algebra (i.e.
non-zero values for the $\{Q,Q\}$ and $\{\bar Q, \bar Q\}$
commutators) comes from a relaxation of the level-matching condition
(\ref{levelmatch}).

\subsection{Restriction to $SU(2|2)$}
\label{sec:rest}

In order to uncover the $SU(2|2)$ structure, we need to decompose the
$SO(8)$ fermions into $SU(2)^4$. This decomposition is discussed in
detail in \cite{Pankiewicz:2003kj}. We note that $\psi^2_\pm$ is in
the ${\bf 8_c}$ of $SO(8)$ while $\psi^1_\pm$ is in the ${\bf 8_s}$.
The decomposition into $SU(2)^4$ is different for the different
$SO(8)$ chiralities
\bsp
&{\bf 8_s} \to ({\bf 2},{\bf 2}) \oplus ({\bf 2'},{\bf 2'}),
~~\text{i.e.}~~\psi^1_a \to {\psi^1_+}_{\a_1\a_2} \oplus {\psi^1_-}^{\da_1\da_2}, \\
&{\bf 8_c} \to ({\bf 2},{\bf 2'}) \oplus ({\bf 2'},{\bf 2}),
~~\text{i.e.}~~\psi^2_{\dot a} \to {\psi^2_+}_{\a_1}^{\da_2} \oplus {\psi^2_-}^{\da_1}_{\a_2} ,\\
\end{split}
\ee  
where $a$ and $\dot a$ run from $1$ to $8$ and
$\a_1,\a_2,\da_1,\da_2$ are the indices of the four $SU(2)$'s. The
indices and gamma matrices are expounded in appendix
\ref{app:gammas}. We are interested in excitations lying in the
$SO(4)$ piece of the geometry (i.e. labelled by indices
$i',j'$). Therefore we restrict our attention to the $X^{i'}$ fields
and their superpartners $\psi^2_-$ and $\psi^1_+$. There is a freedom
in choosing either $Q\sim Q^1_+ + i\g^4 Q^2_-$ or $Q\sim \g^4 Q^1_+ + i
Q^2_-$ for our $SU(2|2)$ supercharge\footnote{Since $Q^1_+$ and $Q^2_-$
  are of different $SO(8)$ chirality, $ Q^1_+ + iQ^2_-$ or  
$ \g^4(Q^1_+ + iQ^2_-)$ mix $SU(2)$ representations and can not
contribute to a $SU(2|2)$ supercharge.}, without loss of generality,
we choose the latter option. Specifically, we define
\bsp
Q^{\da_2}_{\da_1} = \frac{e^{i\pi/4}}{4\pi\a'} \frac{1}{\sqrt{\a'p^+}}
\int_0^{2\pi}d\s
\Biggl\{ &\left(i\p_++\frac{m}{6}\right) X^{\da_2\g_2}
{\psi^2_-}_{\da_1\g_2}\\
+&\left(i\p_- + \frac{m}{6} \right)  X^{\da_2\g_2}\, i
{\s^4}^{\s_1}_{\da_1}\, {\psi^1_+}_{\s_1 \g_2} \Biggr\},\\
\bar Q^{\da_1}_{\da_2} = \frac{e^{-i\pi/4}}{4\pi\a'} \frac{1}{\sqrt{\a'p^+}}
\int_0^{2\pi}d\s
\Biggl\{ &\left(-i\p_++\frac{m}{6}\right) X_{\da_2}^{\g_2}
{\psi^2_-}^{\da_1}_{\g_2}\\
-&\left(-i\p_- + \frac{m}{6} \right)  X_{\da_2}^{\g_2}\, i
{\s^4}^{\s_1\da_1}\, {\psi^1_+}_{\s_1 \g_2} \Biggr\},\\
\end{split}
\ee
where we have defined $X^{\da_2\g_2} = X^{i'} {\s^{i'}}^{\da_2\g_2}$.
We will now express these supercharges in terms of string oscillators.
We will be interested in the action of the algebra on excited states,
and so we leave out the zero-mode part of the following expressions.
The mode expansions for the fields were worked out in detail in
\cite{Hyun:2002wp} and are collected in appendix \ref{app:gammas}.
With these expansions we find
\bsp\label{Qosc}
&Q^{\da_2}_{\da_1} = i\frac{e^{i\pi/4}}{\sqrt{2\a' p^+}} \sum_{n\neq 0}
\O_n \left( \a_n^{\da_2\g_2} {\psi_{-n}}_{\da_1\g_2}+ \tilde
  \a_n^{\da_2\g_2} {\tilde\psi}^4_{-n\,\da_1\g_2} \right),\\
&\bar Q^{\da_1}_{\da_2} = -i\frac{e^{-i\pi/4}}{\sqrt{2\a' p^+}} \sum_{n\neq 0}
\O_{-n} \left( \a_{n\,\da_2}^{\g_2} {\psi_{-n}}_{\g_2}^{\da_1}- \tilde
  \a_{n\,\da_2}^{\g_2} {\tilde\psi}^{4\,\da_1}_{-n\,\g_2} \right),\\
\end{split}
\ee
where ${\tilde\psi}^4_{-n\,\da_1\g_2} =i\, \s^{4\,\a_1}_{\da_1}
{\tilde\psi}_{-n\,\a_1\g_2} $, and where
\be
\O_n =
\frac{1+\frac{6}{m}(\o_n-n)}{\sqrt{1+\left(\frac{6}{m}\right)^2(\o_n-n)^2}},\qquad
\o_n = \sign(n) \sqrt{\left(\frac{m}{6}\right)^2+n^2}.
\ee
In order to accomplish a realization of the $SU(2|2)$ algebra, we must
identify a restricted set of level-I states upon which the algebra
closes. These are states with one oscillator. We choose it to be a
left-moving (untilded) oscillator, but the opposite choice is equally
valid. The main point in uncovering the $SU(2|2)$ structure, as was
discussed previously, is to relax the level-matching condition.  We
therefore do not consider any right-moving excitations.  We define the
(un-level-matched) states
\be\label{levelI}
|\phi^{\db_2}\ra_{\g_2} = \a_{-n\,\g_2}^{\db_2} |0\ra, \qquad
|\psi^{\db_1}\ra_{\g_2} = \psi_{-n\,\g_2}^{\db_1} 
|0\ra,
\ee
where $\g_2$ is a spectator index which we subsequently drop and
$n> 0$. We then find the standard $SU(2|2)$ action
\bsp
&Q^{\da_1}_{\da_2} |\phi^{\db_2}\ra = a\, \d^{\db_2}_{\da_2}
|\psi^{\da_1}\ra,\\
&Q^{\da_1}_{\da_2} |\psi^{\db_1}\ra = b\, \e^{\da_1\db_1}
\e_{\da_2\db_2}|\phi^{\db_2}\ra,\\
&\bar Q^{\da_2}_{\da_1} |\phi^{\db_2}\ra = c\, \e^{\da_2\db_2}
\e_{\da_1\db_1}|\psi^{\db_1}\ra,\\
&\bar Q^{\da_2}_{\da_1} |\psi^{\db_1}\ra = d\, \d^{\db_1}_{\da_1}
|\phi^{\da_2}\ra,\\
\end{split}
\ee
where we have that
\bsp
&a = 2i \frac{e^{i\pi/4}}{\sqrt{2\a'p^+}} \,\O_n\, \o_n, \quad
b = i \frac{e^{i\pi/4}}{\sqrt{2\a'p^+}} \,\O_{-n}, \\
&c = -2i \frac{e^{-i\pi/4}}{\sqrt{2\a'p^+}} \,\O_{-n}\, \o_n, \quad
d =  -i \frac{e^{-i\pi/4}}{\sqrt{2\a'p^+}} \,\O_{n}. \quad
\end{split}
\ee
We note that 
\bsp\label{abcd}
&\frac{1}{2}\left(ad+bc\right) = \frac{\o_n}{\a'p^+},\qquad
ad -bc = \frac{\o_n}{\a'p^+}\left(\O_n^2-\O_{-n}^2\right) \simeq
\frac{\b}{3},\\
&ab = (cd)^*= -\frac{i}{\a'p^+} \O_n \O_{-n} \o_n \simeq  -\frac{in}{\a'p^+},
\end{split}
\ee
where $\simeq$ indicates the $n \ll \a' p^+$ limit. We notice the
consistency with our expectations: the energy of the state (i.e.
$p^-$) is indeed given by $(ad+bc)/2$, while the central charge $ab$
is indeed proportional to the small $n$ limit of $(e^{-in/J}-1)$ with
the consistent proportionality constant $R^2/\a'$ appearing in the
energy\footnote{Here we have used $p^+ = J/R^2$, see (\ref{xpmdef}).}.
Computing the $\{Q,\bar Q\}$ commutator using (\ref{Qosc}), we find
(in the $n \ll \a' p^+$ limit)
\bsp
&\{ Q^{\da_1}_{\da_2},  \bar Q^{\db_2}_{\db_1} \}=
\d_{\db_1}^{\da_1} \d^{\db_2}_{\da_2} H + \frac{\b}{3}\,\d^{\db_2}_{\da_2} {\cal L}_{\db_1}^{\da_1} 
+\frac{\b}{3}\,\d_{\db_1}^{\da_1} {\cal R}^{\db_2}_{\da_2},\\
&\{ Q^{\da_1}_{\da_2},  Q^{\db_1}_{\db_2} \}=
\e^{\da_1\db_1}\e_{\da_2\db_2} {\cal P},\qquad
\{\bar  Q^{\da_2}_{\da_1},  \bar Q^{\db_2}_{\db_1} \}=
\e^{\da_2\db_2}\e_{\da_1\db_1} {\cal K},
\end{split}
\ee
where\footnote{Supersymmetry ensures that the normal ordering
  constants in $H$ and ${\cal P}$ are zero.} 
\bsp
&{\cal R}^{\db_2}_{\da_2} = \frac{1}{2}\sum_{n> 0} \frac{1}{\o_n}
\left(\a^{\db_2\g_2}_{-n} \a_{n\,\da_2\g_2} -
\frac{1}{2}\d^{\db_2}_{\da_2}\,
\a^{\dg_2\g_2}_{-n} \a_{n\,\dg_2\g_2}+\text{right-movers}\right),\\
&{\cal L}_{\db_1}^{\da_1}  = \sum_{n> 0} 
\left( \psi_{-n\,\g_2}^{\da_1}\psi_{n\,\db_1}^{\g_2}
  -\frac{1}{2}\d^{\db_1}_{\da_1}
\psi_{-n\,\g_2}^{\dg_1}\psi_{n\,\dg_1}^{\g_2}+\text{right-movers}\right),\\
&H = \frac{1}{\a'p^+} \sum_{n> 0} \left( \a^{i'}_{-n}  \a^{i'}_{n}
+\o_n \, \psi^{\dg_1}_{-n\,\g_2}\psi_{n\,\dg_1}^{\,\g_2}+\text{right-movers}\right),\\
&{\cal P} = -{\cal K}= -\frac{i}{\a'p^+} \sum_{n> 0}n \left(\frac{1}{\o_n} \a^{i'}_{-n}  \a^{i'}_{n}
+ \psi^{\dg_1}_{-n\,\g_2}\psi_{n\,\dg_1}^{\,\g_2} -\text{right-movers}\right).\\
\end{split}
\ee
Note that ${\cal P}$ is nothing but the level-matching operator
(restricted to the $X^{i'}$ supermultiplet) as previously discussed.
One can then verify that the action of ${\cal R}$ and ${\cal L}$ upon
our states are
\bsp
{\cal R}^{\db_2}_{\da_2}  |\phi^{\dg_2}\ra = 
\d^{\dg_2}_{\da_2} |\phi^{\db_2}\ra - \frac{1}{2} \d^{\db_2}_{\da_2}
|\phi^{\dg_2}\ra,\\
{\cal L}_{\db_1}^{\da_1}  |\psi^{\dg_1}\ra = 
\d^{\dg_1}_{\db_1} |\psi^{\da_1}\ra - \frac{1}{2} \d^{\da_1}_{\db_1}
|\psi^{\dg_1}\ra,\\
\end{split}
\ee
as they should be. Finally we note that the value of $ad-bc$ from
(\ref{abcd}) is what it needs to be in order to close the algebra.  We
have thus found the centrally extended $SU(2|2)$ algebra in the $n \ll
\a' p^+$ limit of the string dual to SYM on $\bR\times S^2$.

\subsection{Generalizations, $S$-matrix, finite-size effects, and giant magnons}
\label{sec:stringgen}

As discussed at the start of section \ref{sec:string}, the IIA
plane-wave appears in the BMN-like limit of the string duals of a rich
class of vacua of any of the three theories: SYM on $\bR\times S^2$,
SYM on $\bR\times S^3/\bZ_k$, or the PWMM.  Thus the $SU(2|2)$ algebra
derived in the last section exists in all of these theories, as long
as the vacuum for the model being studied is well separated. For
analyses similar to those at the start of section \ref{sec:string},
but for $\bR\times S^3/\bZ_k$ and the PWMM, see
\cite{Lin:2005nh}. Having found the $SU(2|2)$ structure, it is natural
to ask whether we can repeat the very rich battery of tests and
analyses which have been carried out in the case of $AdS/CFT$ for
$AdS_5\times S^5$, assuming that our gauge theories really do posses an
all-loop integrable sector. These include the matching of energies of
spinning strings to the thermodynamic limit of the associated
spin-chains, matching the worldsheet S-matrix to gauge theory,
matching finite-size effects (i.e. $1/J$ corrections to the energies
of states), and the existence of solitonic string configurations with
very large worldsheet momentum, the giant magnons. In this section we
will visit each of these issues in a qualitative manner, leaving any
concrete analyses to further work.

\subsubsection{Worldsheet $S$-matrix and finite-size effects}

In order to discuss a worldsheet $S$-matrix, we must have an
interacting sigma model. Since the plane-wave worldsheet theory is
free, one must include curvature corrections in order to develop
worldsheet interactions. The near plane-wave limit is complicated (as
compared to the $AdS_5\times S^5$ case \cite{Callan:2003xr}) by
the dependence of the Lin-Maldacena geometries on the $\r$ and $\eta$
coordinates (the coordinates $r$ and $\theta$ in (\ref{LM})), i.e. the
spatial coordinates transverse to the $S^2$ and $S^5$. The dilaton,
B-field, and Ramond-Ramond field strengths develop dependence on these
coordinates away from the strict plane-wave limit, i.e. their ${\cal
  O}(R^{-2})$ corrections are $\eta$ and $\r$ dependent. However,
despite these complications, if, as we expect, the $SU(2|2)$ symmetry
is exact and so remains at ${\cal O}(R^{-2})$, then the $S$-matrix is
highly constrained by this symmetry, to a single undetermined function
$S^0_{12}$ (see (\ref{S012})) \cite{Beisert:2005tm}. If we consider
the scattering of the bosonic $SO(4)$ excitations (the $X^{i'}$ of
section \ref{sec:LCGPW}) on a subset of states with no excitations
from the $SO(3)$ part of the geometry, then we will find the same
relevant $S$-matrix elements as those found for the $AdS_5\times S^5$
superstring in \cite{Klose:2006zd}. This is the trivial statement that
both theories share an $S^5$ and so share its near plane-wave
geometry. But then the function $S^0_{12}$ is determined at this order
in the large-$R$ expansion. It would therefore not be surprising if
the only change between $AdS_5\times S^5$ and the theories considered
here is that $\l$ is with replaced with $h(\l)$\footnote{Of course
  there will be a different $h(\l)$ for each gauge theory and each
  vacua around which the expansion is being carried out.}, i.e. that
the same expression for the BES phase-factor \cite{Beisert:2006ez}
found in the expression for $S^0_{12}$ in ${\cal N}=4$ SYM is the
relevant one here with $\l\to h(\l)$. Further work would be required
to verify (or disprove) this possibility. Similar statements apply to
the finite-size corrections to the string spectrum. At leading order,
these are given by first-order perturbation theory, and by the same
logic, states built from bosonic $SO(4)$ excitations alone must share
the same leading order finite-size corrections as those found in
$AdS/CFT$. The non-trivial information comes at next-to-leading order,
where second-order perturbation theory must be used, and where $SO(3)$
excitations appear in the intermediate states.

\subsubsection{Spinning strings and giant magnons}

Continuing our analogy with $AdS_5\times S^5$ we may think about
macroscopic spinning strings, corresponding to the thermodynamic limit
of the gauge theory spin-chains. Again, the Lin-Maldacena geometries
contain an $S^5$ which is the site of the $SU(2|2)$ symmetry. Any of
the spinning string solutions with spins only in the $S^5$ may be
borrowed from $AdS_5\times S^5$. The only difference is the modified
relationship between the radius of the $S^5$ and the gauge theory
coupling, i.e. the strong coupling consequence of replacing $\l\to
h(\l)$. The interesting question in this regard are the $1/R^2$
corrections to the energies of these spinning strings. A
semi-classical calculation would require including the fluctuations of
the $SO(3)$ modes, and these have a very different structure than the
corresponding $AdS_5$ modes (c.f. \cite{Tseytlin:2003ii}). It would be
very interesting to attempt such a calculation, which would go towards
fixing $h(\l)$ at next-to-leading order at strong coupling and give
information on the form of the phase-factor in $S^0_{12}$.  The giant
magnon is of course also present in the Lin-Maldacena geometries, for
the same trivial reason that there is an $S^5$ which will accommodate
it. This is consistent with the strong-coupling limit of the $SU(2|2)$
dispersion relation (\ref{HYM}). The finite-size correction to the
energy of the giant magnon \cite{Arutyunov:2006gs,Astolfi:2007uz} does
not require a semi-classical treatment; the calculation takes place
within the $\bR\times S^2$ holding the magnon solution. Thus that
correction is also valid for our case with $\l\to h(\l)$. This
finite-size correction has also been obtained from the integrability
of ${\cal N}=4$ SYM through the Bethe ansatz \cite{Janik:2007wt}.
There it is claimed that the match tests the form of the phase-factor
appearing in $S^0_{12}$. Thus, we have another piece of evidence
suggesting that the $\l\to h(\l)$ replacement may bring us from ${\cal
  N}=4$ SYM, to the potentially integrable sector described here for
SYM on $\bR\times S^2$, SYM on $\bR\times S^3/\bZ_k$, and the PWMM.

\section{Concluding remarks}
\label{sec:final}
In this paper we have found an application for the rich constraining
power of the mass-deformed $SU(2|2)$ algebra in the gauge/gravity
duality for $\mathcal{N}=4 $ SYM on $\mathbb{R}\times S^3$ and its
dimensional reductions. We have mostly focused on the three
dimensional $\mathcal{N}=8 $ SYM on $\mathbb{R}\times S^2$ and its
string dual to illustrate the use of this superalgebra in the
computation of the spectrum of the gauge as well as the dual
world-sheet theory.  The two-loop gauge theory results and the leading
order strong coupling computations done in the plane wave limit of the
associated string theory suggest a potentially integrable $SU(2|3)$
sector for this particular realization of the gauge/gravity duality.
Moreover, we find that various quantities, such as the form of the
all-loop dispersion relation and the ``matrix'' structure of the
S-matrix for the gauge theory Hamiltonian are exactly the same in this
theory as the planar dilatation operator for $\mathcal{N}=4$ SYM. 

Despite the methodological similarities, there are various fundamental
differences that distinguish the three and four dimensional gauge
theories from each other. For instance, analytical dependence of the
physical spectrum on the effective 't Hooft couplings, encoded in the
function $h(\lambda)$. Furthermore, the world sheet theory for the
three dimensional gauge theory is not a coset model which makes the
issue of understanding even its classical integrability challenging.
Finally, the gauge theory appears to possess multiple vacuaa (which is
reflected on the string theory side in the various disc configurations
discussed earlier), which is unlike the physical behavior of the four
dimensional superconformal theory.  It is thus gratifying that despite
these differences, various physical quantities can be analyzed in both
these gauge theories using similar techniques of analysis. Apart from
another venue for the potential utilization of the powerful algebraic
methods tied to integrable structures, our study also opens up some
interesting lines of investigation which we comment on below.

An obvious question to ask is whether or not the tell-tale signs of
integrability for the three dimensional theory translate into all-loop
integrability. Integrability to all orders, even in a restricted
sub-sector of the theory (like the $SU(2)$ sector) would be a powerful
boost towards performing a comprehensive test for the gauge/gravity
duality without the use of conformal symmetries.  Assuming
integrability holds in a subsector of the gauge theory, the complete
determination of the interpolating $h$ function, which we have
computed at weak and strong coupling, is certainly an extremely
interesting question and might be amenable to analysis by methods
such as the Y-system, which has recently yielded dramatic results
\cite{Gromov:2009tv}.

As mentioned before in the paper, a fuller understanding of the the
gauge/gravity duality for the other dimensional reductions of the
four dimensional gauge theory might be gained by adapting the
present algebraic techniques accordingly. In particular, the
interplay between the non-trivial vacua for the gauge theories in
question and mass-deformed algebras is another potential line of
investigation coming out of the present analysis.

Looking beyond the immediate concerns of this paper and the
computation of spectra; the r\^ole of supersymmetry in the study of
other extended degrees of freedom, such as Wilson loops would be
another interesting problem to study. The massless version of the
three dimensional theory defined on $\mathbb{R}^3$ has recently been
shown to posses a large class of BPS Wilson loops whose expectation
values are completely determined by supersymmetry
\cite{Agarwal:2009up}.  The investigation of the corresponding
extended objects of the massive theory on $\mathbb{R}\times S^2$
should yield further valuable information for the field theory and its
string theoretic counterpart.

On a final, somewhat tangential note, it is worth pointing out that
massive algebras also arise in the context of three dimensional SYM
theories even in flat backgrounds with minimal supersymmetry
\cite{Agarwal:2009gb}.  In these theories the supersymmetry algebra is
deformed by the spacetime rotation group as opposed to the
$R$-symmetry group. Furthermore, the mass-gap for the gluonic fields
in these theories is generated by a term that is closely related to
the volume measure on the gauge invariant configuration space for pure
Yang-Mills theory in three dimensions \cite{Karabali:1997wk} (for recent progress in three dimensional pure Yang-Mill theory on $\mathbb{R}\times S^2$, see \cite{Agarwal:2008dk}).  This is
perhaps indicative of a potential deeper connection between massive
supersymmetry algebras and dynamical mass-generation in confining
three dimensional SYM theories.

\section{Acknowledgements}

We would like to thank Niklas Beisert and Jan Plefka for discussions
and their comments on a previous version of this manuscript. DY is
supported by the Volkswagen Foundation.


\appendix

\section{Gamma matrices and mode expansions}
\label{app:gammas}

The $SO(1,9)$ gamma matrices are given by \cite{Hyun:2002wu}
\bsp
&\G^0 = -i \s^2 \otimes {\bf 1}_{16},\quad
\G^{11} = \s^1 \otimes {\bf 1}_{16},\quad
\G^I = \s^3 \otimes \g^I,\\
&\G^9 = -\s^3 \otimes \g^9,\quad
\G^\pm = \frac{1}{\sqrt{2}} ( \G^0 \pm \G^{11}),
\end{split}
\ee
where $\g^9 = \g^{12345678}$, where $\g^I$ are the 16$\times$16
$SO(8)$ gamma matrices. We choose the following representation for
them
\be\label{so8g}
\g^I = \begin{pmatrix}
0~~~ \tilde \g^I_{a\dot a}\\
\tilde\g^I_{\dot a a}~~~ 0
\end{pmatrix},
\ee
where $a,\dot a$ run from 1 to 8. We decompose the $\tilde\g^I$ into
two $SU(2)\times SU(2)$ representations (one for each $SO(4)$):
\bea 
&&\tilde\gamma^i_{a\dot{a}} = \left(\begin{matrix} 0 &
\s^i_{\a_1\db_1}\d_{\a_2}^{\b_2} \cr
{\s^i}^{\da_1\b_1}\d^{\da_2}_{\db_2} & 0\end{matrix} \right)\
,\qquad~~ \tilde\g^i_{\dot{a}a} =\left(
\begin{matrix}
0 & \s^i_{\a_1\db_1}\d^{\da_2}_{\db_2} \cr
{\s^i}^{\da_1\b_1}\d_{\a_2}^{\b_2} & 0\end{matrix} \right)\ ,\\
&&\tilde\g^{i'}_{a\dot{a}} =\left(
\begin{matrix}
-\d_{\a_1}^{\b_1}\s^{i'}_{\a_2\db_2} & 0 \cr0 &
\d^{\da_1}_{\db_1}{\s^{i'}}^{\da_2\b_2}\end{matrix} \right)\ ,\qquad~~
 \tilde\g^{i'}_{\dot{a}a} =\left( 
\begin{matrix}
-\d_{\a_1}^{\b_1}{\s^{i'}}^{\da_2\b_2} & 0 \cr 0 &
\d^{\da_1}_{\db_1}\s^{i'}_{\a_2\db_2}\end{matrix} \right)\ . \eea 
The $SU(2)$ indices $\a_i,\b_i$ are raised and lowered using
$\e_{\a\b}$, where $\e_{12} = -\e_{21} =1$, and similarly for the
dotted indices. The $\s$-matrices satisfy the relations
\begin{equation}
\s^i_{\a\da}{\s^j}^{\da\b}+\s^j_{\a\da}{\s^i}^{\da\b}
=2\d^{ij}\d_{\a}^{\b}\,,\qquad
{\s^i}^{\da\a}\s^j_{\a\db}+{\s^j}^{\da\a}\s^i_{\a\db}=
2\d^{ij}\d^{\da}_{\db}\,.
\end{equation}
Some other properties satisfied by these matrices are
\bea
\label{rel1}
&&\e_{\a\b}\e^{\g\d}  = \d_{\a}^{\d}\d_{\b}^{\g}-\d_{\a}^{\g}\d_{\b}^{\d}\,,\\
&&\s^i_{\a\db}{\s^j}^{\db}_{\b}  = -\d^{ij}\e_{\a\b}+\s^{ij}_{\a\b}\,,\qquad
(\s^{ij}_{\a\b}\equiv \s^{[i}_{\a\da}{\s^{j]}}^{\da}_{\b}=\s^{ij}_{\b\a})\\
&&\s^i_{\a\da}{\s^j}^{\a}_{\db} = -\d^{ij}\e_{\da\db}+\s^{ij}_{\da\db}\,,
\qquad (\s^{ij}_{\da\db}\equiv \s^{[i}_{\a\da}{\s^{j]}}^{\a}_{\db}=\s^{ij}_{\db\da})\\
&&\s^k_{\a\da}\s^{k}_{\b\db}  = 2\e_{\a\b}\e_{\da\db}\,,\\
&&\s^{kl}_{\a\b}\s^{kl}_{\g\d}  = 4(\e_{\a\g}\e_{\b\d}+\e_{\a\d}\e_{\b\g})\,,\\
&&\s^{kl}_{\a\b}\s^{kl}_{\dg\dd}  = 0\,,\\
\label{rel7}
&&2\s^i_{\a\da}\s^{j}_{\b\db} = \d^{ij}\e_{\a\b}\e_{\da\db}
+\s^{k(i}_{\a_1\b_1}\s^{j)k}_{\da_1\db_1}
-\e_{\a\b}\s^{ij}_{\da\db}-\s^{ij}_{\a\b}\e_{\da\db}\,.
\eea

The following Fierz identities are also useful for calculating the
commutator of the supercharges,
\be
\left(\psi^A_{\pm}\right)_\g \left(\psi^A_{\pm}\right)_\d =
\frac{1}{4}\d_{\g\d}\psi^A_\pm\psi^A_\pm
  + \frac{1}{8} \g^{ij}_{\g\d}\,\psi^A_\pm \g^{ij}
\psi^A_\pm  + \frac{1}{8} \g^{i'j'}_{\g\d}\,\psi^A_\pm \g^{i'j'}
\psi^A_\pm ,
\ee
\bsp
\left(\psi^A_{\pm}\right)_\g \left(\psi^B_{\mp}\right)_\d =
&\frac{1}{4}\g^i_{\g\d}\,\psi^A_\pm\g^i \psi^B_\mp
+\frac{1}{4}\g^{i'}_{\g\d}\,\psi^A_\pm\g^{i'} \psi^B_\mp,\\
\end{split}
\ee
where no summation is implied on the $A,B$ indices and $\g,\d$ are
$SO(8)$ spinor indices running from 1 to 16. 

The mode expansions (excluding zero-modes), for the string embedding
functions are given by \cite{Hyun:2002wp}
\bsp
&X^{i'} = i\sqrt{\frac{\a'}{2}} \sum_{n\neq 0} \frac{1}{\o_n} \left(
  \a_n^{i'} \phi_n^* + \tilde \a_n^{i'} \phi_n \right),\quad
\dot X^{i'} = \sqrt{\frac{\a'}{2}} \sum_{n\neq 0} \left(
  \a_n^{i'} \phi_n^* + \tilde \a_n^{i'} \phi_n \right),\\
&{\psi^2_-}^{\da_1}_{\a_2} = \sum_{n\neq 0} c_n
\left( {\psi_n}^{\da_1}_{\a_2}\, \phi_n^* -i \frac{6}{m}(\o_n - n)
{\s^4}^{\da_1\g_1} {\tilde\psi}_{n\,\g_1\a_2} \,\phi_n \right),\\
&{\psi^1_+}_{\a_1\a_2} = \sum_{n\neq 0} c_n
\left( {\psi_n}_{\a_1\a_2}\, \phi_n^* +i \frac{6}{m}(\o_n - n)
{\s^4}_{\a_1\dg_1} {\tilde\psi}^{~\,\dg_1}_{n\,\a_2} \,\phi_n \right),\\
\end{split}
\ee
where
\bsp
&\o_n = \sign(n) \sqrt{\left(\frac{m}{6}\right)^2+n^2},\quad
\phi_n = e^{in\s},\quad c_n =
\frac{\sqrt{\a'}}{\sqrt{1+\left(\frac{6}{m}\right)^2(\o_n-n)^2}},\\
&\{{\psi_n}^{\da_1}_{\a_2},  {\psi_m}^{\b_2}_{\db_1}\} = \d_{n+m} 
\d^{\da_1}_{\db_1} \d^{\b_2}_{\a_2}, \quad
\{{\tilde \psi}^{\,~\a_1}_{n\,\a_2},  {\tilde \psi}^{~~\b_2}_{m\,\b_1}\} = \d_{n+m} 
\d^{\a_1}_{\b_1} \d^{\b_2}_{\a_2},\\
&[\a_n^{i'},\a_m^{j'}] = \o_n \d_{n+m} \d^{i'j'},\quad
[\tilde \a_n^{i'},\tilde \a_m^{j'}] = \o_n \d_{n+m} \d^{i'j'},
\end{split}
\ee
and so negative mode numbers represent creation operators. 

\section{Relation between $SO(6)$ and $SU(4)$}
\label{sec:so6su4}

The $SO(9,1)$ gamma matrices are given by
\be
\G^{\mu}=\g^{\mu}\otimes 1_8,\quad 
\G^{AB}=\g_5\otimes\begin{pmatrix}0&-\tilde{\rho}^{AB}\\ \rho^{AB}&0\end{pmatrix}
=-\G^{BA}.
\ee
$\G^{AB}$ satisfies $\{\G^{AB},\G^{CD}\}=\e^{ABCD}$.  $\rho^{AB}$ and
$\tilde{\rho}^{AB}$ are defined by
\be
(\rho^{AB})_{CD}=\d^A_C\d^B_D-\d^A_D\d^B_C,\quad
(\tilde{\rho}^{AB})^{CD}=\frac{1}{2}\e^{CDEF}(\rho^{AB})_{EF}=\e^{ABCD}.
\ee
The relationship between the $SO(6)$ and $SU(4)$ bases is
\bsp
&X^{AB}=\frac{1}{2}\e^{ABCD}X_{CD},\quad X^{AB}=-X^{BA}=X_{AB}^\dag,\quad
X_{i4}=\frac{1}{2}(X_{i}+iX_{i+3}) \\
&\G^{i4}=\frac{1}{2}(\G^{i}-i\G^{i+3}),\quad X_{AB}\G^{AB}=X_m\G^m.
\end{split}
\ee
%

\bibliography{glueball}
\end{document}